\documentclass[12pt,preprint]{aastex}

\newcommand\iso[2]{$^{\rm #1}$#2}

\shorttitle{$\omega$ Centauri Abundances}
\shortauthors{Johnson et al.}

\begin{document}

\title{Fe and Al Abundances for 180 Red Giants in the Globular Cluster Omega
Centauri (NGC 5139)}

\author{Christian I. Johnson\altaffilmark{1}, 
Catherine A. Pilachowski\altaffilmark{1}, Jennifer Simmerer\altaffilmark{2},
and Dustin Schwenk\altaffilmark{3}
}

\altaffiltext{1}{Department of Astronomy, Indiana University,
Swain West 319, 727 East Third Street, Bloomington, IN 47405--7105, USA;
cijohnson@astro.indiana.edu; catyp@astro.indiana.edu}

\altaffiltext{2}{Lund Observatory, Box 43, SE 221-00 Lund, Sweden; 
jennifer@astro.lu.se}

\altaffiltext{3}{Department of Physics, University of Illinois at 
Urbana--Champaign, 2910 Artesia Crossing, Urbana, IL 61802, USA; 
schwenk@uiuc.edu}

\begin{abstract}

We present radial velocities, Fe, and Al abundances for 180 red giant branch
(RGB) stars in the Galactic globular cluster Omega Centauri ($\omega$ Cen).
The majority of our data lie in the range 11.0$<$V$<$13.5, which covers the RGB
from about 1 mag. above the horizontal branch to the RGB tip.  The selection
procedures are biased towards preferentially observing the more metal--poor
and luminous stars of $\omega$ Cen.  Abundances 
were determined using equivalent width measurements and spectrum synthesis
analyses of moderate resolution spectra (R$\approx$13,000) obtained with the
Blanco 4m telescope and Hydra multifiber spectrograph.  Our results are in
agreement with previous studies as we find at least four different metallicity
populations with [Fe/H]=--1.75, --1.45, --1.05, and --0.75, with a full range
of --2.20$\la$[Fe/H]$\la$--0.70.  [Al/Fe] ratios exhibit large star--to--star
scatter for all populations, with the more than 1.0 dex range of [Al/Fe] 
decreasing for stars more metal--rich than [Fe/H]$\sim$--1.4.  The minimum 
[Al/Fe] abundance observed for all metallicity populations is 
[Al/Fe]$\sim$+0.15.  The maximum abundance of log $\epsilon$(Al) is reached 
for stars with [Fe/H]$\sim$--1.4 and does not increase further with stellar 
metallicity.  We interpret these results as evidence for type II SNe 
providing the minimum [Al/Fe] ratio and a mass spectrum of intermediate mass 
asymptotic giant branch stars causing the majority of the [Al/Fe] scatter.
These results seem to fit in the adopted scheme that star formation occurred 
in $\omega$ Cen over $>$1 Gyr.

\end{abstract}


\section{INTRODUCTION}

The Galactic globular cluster Omega Centauri ($\omega$ Cen) presents a unique
opportunity to study the chemical evolution of both a small stellar system
and stars with common formation histories covering a metallicity range of more 
than a factor of 10, a defining characteristic of $\omega$ Cen that has been
known since the initial discovery of its unusually broad red giant branch (RGB) 
by Woolley (1966).  Although $\omega$ Cen is the most massive Galactic globular
cluster, with an estimated mass of $\sim$2--7$\times$10$^{\rm 6}$ 
M$_{\sun}$ (Richer et al. 1991; Meylan et al. 1995; van de Ven et al. 2006),
it does not appear to have an exceptionally deep gravitational potential
well (Gnedin et al. 2002).  This seems to negate a simple explanation that
$\omega$ Cen evolved as a typical globular cluster that was more easily able 
to retain supernova (SN) and asymptotic giant branch (AGB) ejecta for 
self--enrichment.  This fact coupled with the cluster's retrograde 
orbit and disk crossing time of $\sim$1--2$\times$10$^{\rm 8}$ years (e.g., 
Dinescu et al. 1999), which could severely inhibit star formation, are some of 
the strongest arguments against $\omega$ Cen having a Galactic origin.  
Instead, it has been proposed (e.g., Dinescu et al. 1999; Smith et al. 2000; 
Gnedin et al. 2002; Bekki \& Norris 2006) that $\omega$ Cen may be the 
remaining nucleus of a dwarf spheroidal galaxy that evolved in isolation and 
was later accreted by the Milky Way, suggesting the progenitor system was 
perhaps a factor of 100--1000 times more massive than what is presently 
observed.

Recent spectroscopic and photometric studies (Norris \& Da Costa 1995; Norris 
et al. 1996; Suntzeff \& Kraft 1996; Lee et al. 1999; Hilker \& Richtler 2000; 
Hughes \& Wallerstein 2000; Pancino et al. 2000; Smith et al. 2000; van 
Leeuwen et al. 2000; Rey et al. 2004; Stanford et al. 2004; 
Piotto et al. 2005; Sollima et al. 2005a; Sollima et al. 2005b; Kayser et al. 
2006; Sollima et al. 2006; Stanford et al. 2006; Stanford et al. 2007; van Loon
et al. 2007; Villanova et al. 2007) have confirmed the existence of up to five 
separate stellar populations ranging in metallicity from [Fe/H]$\sim$--2.2 to 
--0.5, with a peak in the metallicity distribution near [Fe/H]$\sim$--1.7 and 
a long tail extending to higher metallicities.  In addition to the metal--poor 
and intermediate metallicity populations initially seen in the Woolley 
(1966) photometric study, Lee et al. (1999) and Pancino et al. (2000) 
discovered the existence of the most metal--rich RGB at [Fe/H]$\sim$--0.5,
commonly referred to as the anomalous RGB (RGB--a).  The RGB--a is primarily 
observed in the central region of the cluster and contains approximately 5$\%$ 
of the total stellar population (Pancino et al. 2000), in contrast to the 
dominant metal--poor population that contains roughly 75$\%$ of cluster stars.
Additionally, there is some evidence (Norris et al. 1997) that the metal--rich
population exhibits smaller radial velocity dispersion and rotation than the 
metal--poor population.  Sollima et al. (2005b) confirmed the Norris et al. 
(1997) results but also showed that the most metal--rich stars ([Fe/H]$>$--1)
exhibit an increasing velocity dispersion as a function of increasing 
metallicity, which could be evidence for accretion events occurring within
$\omega$ Cen's progenitor system (Ferraro et al. 2002; Pancino et al. 2003);
however, this result is not yet confirmed (Platais et al. 2003, but see also
Hughes et al. 2004).  It should be noted that Pancino et al. (2007), using
radial velocity measurements of 650 members with measurement uncertainties of 
order 0.5 km s$^{\rm -1}$, have found no evidence for rotational differences
among the different metallicity groups.
 
The distribution of main--sequence turnoff (MSTO) and subgiant branch (SGB) 
stars matches that observed on the RGB, such that one can trace the 
evolutionary sequence of each population from at least the MSTO to the RGB 
using high precision photometry (e.g., Villanova et al. 2007).  The
main--sequence (MS) has proved equally as complex as the SGB and RGB, with the 
discovery by Anderson (1997) of a red and blue MS (BMS).  Interestingly, 
Piotto et al. (2005) discovered that the BMS was more metal--rich than the 
red MS, suggesting the BMS could be explained assuming a higher He content, 
perhaps as high as Y$\sim$0.38 (Bedin et al. 2004; Norris 2004; Lee et al. 
2005; Piotto et al. 2005).

While it is clear that multiple populations are present in this 
cluster, there has been some debate regarding the age of each population.  
There is general agreement that the age range is between about 0 and 6 Gyrs 
(Norris \& Da Costa 1995; Hilker \& Richtler 2000; Hughes \& Wallerstein 2000; 
Pancino et al. 2002; Origlia et al. 2003; Ferraro et al. 2004; Hilker et al. 
2004; Rey et al. 2004; Sollima et al. 2005a; Sollima et al. 2005b; Villanova 
et al. 2007), though the recent work by Stanford et al. (2006) suggests the 
most likely age range is $\sim$2--4 Gyrs, with the metal--rich stars being 
younger.  For the case of monotonic chemical enrichment in a single system, 
one would expect the more metal--rich stars to be younger than the more 
metal--poor; however, this assumption has been questioned by Villanova et al. 
(2007) who suggested the metal--rich stars and 33$\%$ of the metal--poor stars 
are the oldest with the remaining 2/3 of the metal--poor population being 3--4 
Gyrs younger.  The picture of $\omega$ Cen's formation is further compounded 
by observations of RR Lyrae horizontal branch (HB) stars that reveal 
a bimodal metallicity distribution \emph{without} a trend in He enhancement as 
a function of [Fe/H] (Sollima et al. 2006).  The important point here is that a 
group of RR Lyrae stars exists with the same metallicity as the BMS but 
without the presumed He enhancement.  A He--rich secondary population would 
not produce a significant RR Lyrae population unless a $\ga$4 Gyr age 
difference was present with respect to the dominant metal--poor population 
(Sollima et al. 2006).  The required age difference is therefore inconsistent 
with most age spread estimates that put $\Delta$$\tau$$\la$4 Gyrs.

$\omega$ Cen's chemical evolution history has so far proved difficult to
interpret from measured abundances of light (Z$\la$27), $\alpha$, Fe--peak, 
s--process, and r--process elements.  In ``normal" Galactic globular clusters,
C, N, O, F, Na, Mg (sometimes), and Al often exhibit large star--to--star 
variations, in some cases exceeding more than a factor of 10 (e.g., see recent 
review by Gratton et al. 2004).  In contrast, the heavier $\alpha$--elements 
(e.g., Ca and Ti) show little to no variation and are enhanced relative to Fe
at [$\alpha$/Fe]$\sim$+0.30, with a decreasing ratio for clusters with
[Fe/H]$>$--1.  Likewise, Fe and all other Fe--peak, s--process, and r--process
elements show star--to--star variations of $\sim$0.10--0.30 dex.  
Additionally, nearly all globular clusters are enriched in r--process 
relative to s--process elements by about 0.20 dex.  In $\omega$ Cen, [Fe/H]
covers a range of more than 1.5 dex and, as previously stated, it has a 
potential well comparable to that of other globular clusters, suggesting it 
had to be different in the past to undergo self--enrichment.  The scenario of 
two or more globular clusters merging seems unlikely now given the results of 
Pancino et al. (2007) and the typically large orbital velocities coupled with 
the small velocity dispersions of clusters (Ikuta \& Arimoto 2000).  While 
$\omega$ Cen exhibits large abundance variations for several of the light
elements at various metallicities (e.g., Norris \& Da Costa 1995; Smith et al.
2000), the mean heavy $\alpha$--element enhancement is surprisingly uniform at 
[$\alpha$/Fe]$\sim$+0.30 to +0.50 (Norris \& Da Costa 1995; Smith et al. 2000;
Villanova et al. 2007), with perhaps a trend of decreasing [$\alpha$/Fe] at 
[Fe/H]$>$--1 (Pancino et al. 2002).  The s--process elements show a 
clear increase in abundance relative to Fe with a plateau occurring at 
[Fe/H]$\sim$--1.40 to --1.20 (Norris \& Da Costa 1995; Smith et al. 2000).  
However, unlike in globular clusters, s--process elements are overabundant
with respect to r--process elements, where [Ba/Eu] typically reaches between
0.5 and 1.0 (Smith et al. 2000), indicating a strong presence of AGB ejecta.

Many globular cluster giants show clear C--N, O--Na, O--Al, Mg--Al, and in the 
case of M4 (Smith et al. 2005), F--Na anticorrelations alongside a Na--Al 
correlation (e.g., Gratton et al. 2004).  In addition to these anomalies being
present in the atmospheres of RGB stars, similar relations have been observed
in some globular cluster MS and MSTO stars (e.g., Cannon et al. 1998; 
Gratton et al. 2001; Cohen et al. 2002; Briley et al. 2004a; 2004b; Boesgaard 
et al. 2005).  According to standard evolutionary theory, first dredgeup 
brings the products of MS core hydrogen burning to the surface and homogenizes 
approximately 70--80$\%$ of the star, resulting in C depletion, N enhancement, 
and a lowering of the \iso{12}{C}/\iso{13}{C} ratio from about 90 to 25 (e.g., 
Salaris et al. 2002).  The decline in [C/Fe] and \iso{12}{C}/\iso{13}{C} has 
been verified via observations in both globular cluster (Bell et al. 1979; 
Carbon et al. 1982; Langer et al. 1986; Bellman et al. 2001) and field stars 
(Charbonnel \& do Nascimento 1998; Gratton et al. 2000; Keller et al. 2001) as 
strong evidence for in situ mixing occurring along the RGB.  However, as the 
advancing hydrogen--burning shell (HBS) crosses the molecular weight 
discontinuity left by the convective envelope's deepest point of penetration, 
extra mixing not predicted by canonical theory occurs in both
field and cluster stars, driving down [C/Fe] further and allowing 
\iso{12}{C}/\iso{13}{C} to reach the CN--cycle equilibrium value of $\sim$4.
The mechanism responsible for this extra mixing is not known, though 
thermohaline mixing (Charbonnel \& Zahn 2007) may ameliorate the problem.
While halo field and cluster giants share these same trends, differences arise
when considering O, Na, and Al abundances.  Field stars do not exhibit most of 
the familiar correlations/anticorrelations and large star--to--star variations 
seen in globular cluster stars and instead remain mostly constant from the MS 
to the RGB tip (e.g., Ryan et al. 1996; Fulbright 2000; Gratton et al. 2000).

The reason for the observed differences between cluster and field giants is 
not known, but obviously the higher density cluster environment is a key 
factor.  Coupled O depletions and Na/Al enhancements are clear signs of
high temperature (T$\ga$40$\times$10$^{\rm 6}$ K) H--burning via the ON, NeNa, 
and MgAl proton--capture cycles, but this does not necessarily mean those cycles
are operating in the RGB stars we presently observe and instead may be from the
ejecta of intermediate mass (IM) AGB stars ($\sim$3--8 M$_{\sun}$) that 
underwent hot bottom burning (HBB) and polluted the gas from which the current
stars formed.  One of the strongest arguments against in 
situ mixing is the observed abundance relations on the MS and MSTO matching 
those on the RGB because these stars are both too cool for the ON, NeNa, and 
MgAl cycles to operate and their shallow envelope convection zones do not 
reach deep enough to bring up even CN--cycled material.  Additionally, 
Shetrone (1996) showed that at least in M13 giants, \iso{24}{Mg} is 
anticorrelated with Al instead of \iso{25}{Mg} and/or \iso{26}{Mg}, which 
means temperatures not achievable in low mass RGB stars (at least 
70$\times$10$^{\rm 6}$ K) are needed to activate the full MgAl chain (Langer 
et al. 1997); however, these temperatures are reached in HBB conditions.  
Current models of low mass RGB stars (e.g., Denissenkov \& Weiss 2001) 
indicate \iso{27}{Al} is only produced deep in the stellar interior by 
burning \iso{25}{Mg} and convective mixing reaching these depths would cause 
a second increase in the surface abundance of both \iso{23}{Na} and 
\iso{4}{He}.  It should be noted that if it is instead \iso{26}{Al} 
($\tau$$_{\rm 1/2}$$\sim$1$\times$10$^{\rm 6}$ yrs) causing the abundance 
anomalies on the upper RGB, then the O--Na and Na--Al relations can be 
explained in a self--consistent manner via in situ mixing (Denissenkov \& 
Weiss 2001).  Also, there is some evidence that O depletions and Na/Al 
enhancements become stronger in the upper $\sim$0.7 mag before the RGB tip in 
M13 (e.g., Sneden et al. 2004; Johnson et al. 2005), indicating the possible 
operation of additional deep mixing episodes in some stars.  Although it is 
more difficult to believe in situ mixing is responsible for the 
\iso{24}{Mg}--\iso{27}{Al} anticorrelation, the same may not be true for O and 
Na.  In or just above the HBS of a metal--poor low mass RGB star, the O--Na 
anticorrelation can be naturally explained because the ON and NeNa cycles can 
operate at T$\sim$40$\times$10$^{\rm 6}$ K (Denisenkov \& Denisenkova 1990; 
Langer et al. 1993).  Of course, this cannot be the case for any O--Na 
anticorrelation observed in MSTO and SGB stars and does require convective
mixing in RGB stars to penetrate past the radiative zone separating the bottom 
of the convective envelope and the top of the HBS.

While pollution from a previous generation of more massive AGB stars seems an 
attractive explanation, there are a few important issues.  Predicted IM--AGB 
stellar yields are sensitive to the adopted treatment of convection because it 
affects other important parameters such as luminosity, number of thermal 
pulses, third dredgeup efficiency, envelope temperature structure, and mass
loss (Ventura \& D'Antona 2005a).  The two most common methods employed are
mixing length theory (MLT) (e.g., Fenner et al. 2004) and the full spectrum of 
turbulence (FST) model (e.g., Ventura \& D'Antona 2005b), with the latter 
providing more efficient convection.  In $\omega$ Cen and all other globular 
clusters observed, the [C+N+O/Fe] sum is constant (Pilachowski et al. 1988; 
Dickens et al. 1991; Norris \& Da Costa 1995; Smith et al. 1996; Ivans et al. 
1999), but models based on MLT indicate stars forming from different 
generations of AGB ejecta should show a large increase in the CNO sum (e.g., 
Lattanzio et al. 2004).  In contrast, FST models keep [C+N+O/Fe] constant to 
within about a factor of 2 due to enhanced mass loss and fewer third dredgeup
episodes (Ventura \& D'Antona 2005b).  Although Na and Al production could be 
due to HBB, it is difficult to produce the observed O depletion of 1.0 to 1.5 
dex along with the required Na enhancement (e.g., Denissenkov \& Herwig 2003; 
but see also Ventura \& D'Antona 2005b).  Self--consistent models
of globular cluster enrichment from AGB ejecta fail to reproduce the MgAl
anticorrelation seen in several globular clusters, including $\omega$ Cen, where
Mg increases relative to Al instead of decreases (Fenner et al. 2004).
Without an evolutionary scenario, O deficient, Na/Al enhanced stars 
must have preferentially formed out of enriched gas relative to ``O--normal"
stars (i.e., [O/Fe]$\sim$+0.30) and Yong et al. (2003) point out that 
even with no O present in the enriched gas, these stars would require
a composition of 90$\%$ enriched, 10$\%$ ``normal" material to obtain the
observed O deficiency.  Lastly, AGB stellar envelopes contain roughly 36$\%$
He by mass (Lattanzio et al. 2004), but O--poor, Na/Al--rich stars do not
appear to be particularly He--rich; however, this does not rule out AGB stars
as the source of the He--rich BMS observed in $\omega$ Cen.  Given the 
evidence for and against evolutionary and primordial processes, a hybrid 
scenario probably needs to be invoked to explain all abundance anomalies.

Given the inherently large spread in metallicity of stars in $\omega$ Cen and 
that Al is the heaviest element sensitive to proton--capture 
nucleosynthesis at temperatures achieved in the interiors of low mass 
metal--poor RGB stars, we present radial velocities, Fe, and Al abundances for 
180 RGB stars covering --2.20$<$[Fe/H]$<$--0.70.  With additional data from
the literature covering from the MS to the RGB tip, we address the issues of 
star formation and possible pollution sources driving the chemical evolution
of $\omega$ Cen as a function of metallicity.
 
\section{OBSERVATIONS AND REDUCTIONS}

The observations of all 180 giants in $\omega$ Cen were obtained with the Blanco
4m telescope using the Hydra multifiber positioner and bench spectrograph at 
the Cerro Tololo Inter--American Observatory.  All observations were obtained 
using the ``large" 300$\micron$ (2$\arcsec$) fibers.  The full spectral 
coverage ranged from $\sim$6450--6750~\AA, centered on $\sim$6600~\AA;
however, wavelengths blueward of $\sim$6500~\AA\ lie on the shoulder of the 
filter response curve, making continuum placement difficult.  Therefore, we 
truncated the spectra to include only the region from 6500--6750~\AA.  The 
316 line mm$^{\rm -1}$ echelle grating and Blue Air Schmidt Camera provided
a resolving power of R($\lambda$/$\Delta$$\lambda$)$\approx$13,000 (0.5~\AA\ 
FWHM) at 6600~\AA.  A list of our observation dates and exposure times is 
provided in Table 1.

\begin{figure}
\epsscale{0.90}
\plotone{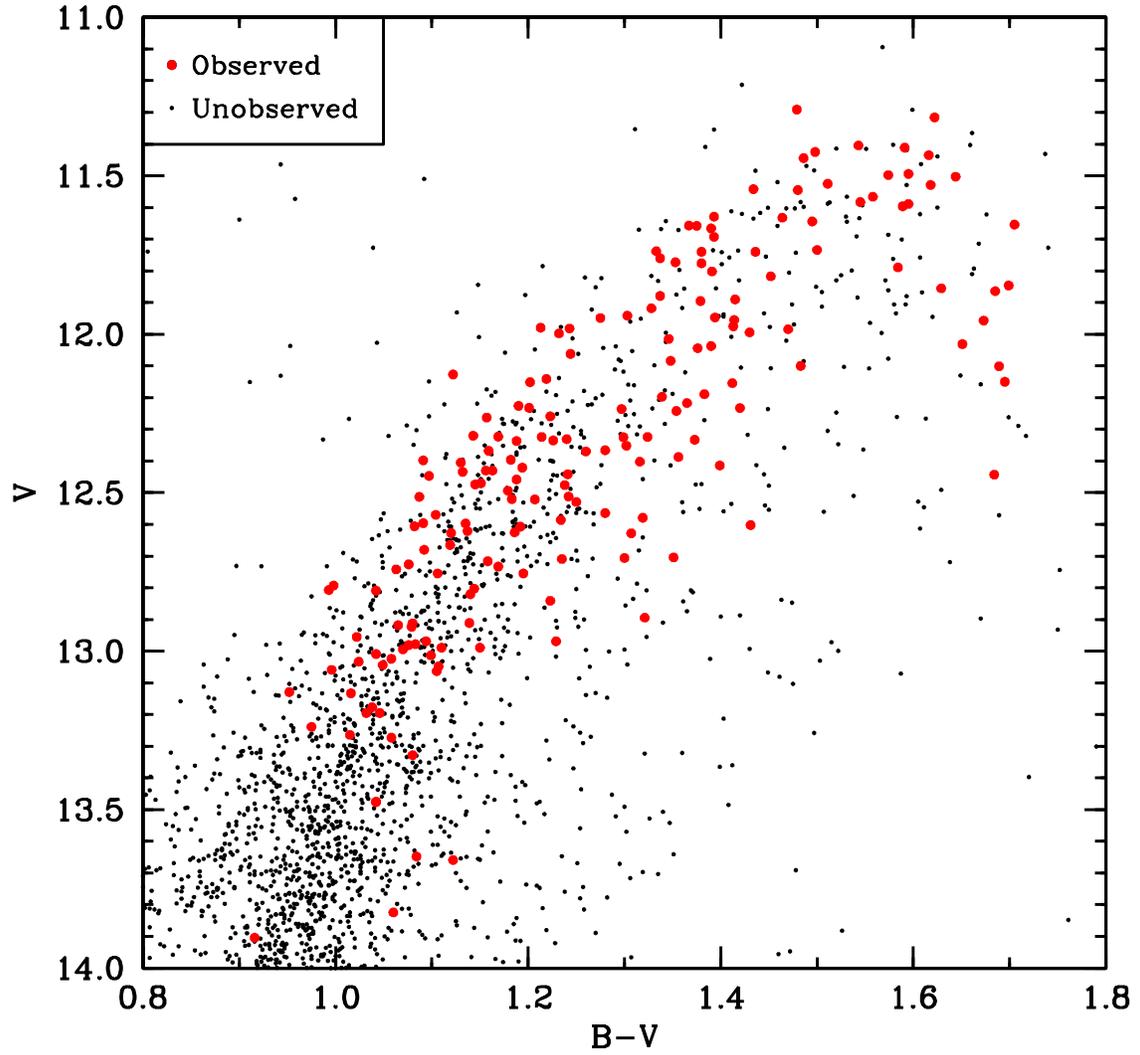}
\caption{A color--magnitude diagram of the upper RGB for $\omega$ Cen.  The 
large filled circles indicate program stars and the small filled circles
are those available from the van Leeuwen et al. (2000) proper motion study.}
\label{f1}
\end{figure}

Target stars, coordinates, photometry, and membership probability were taken 
from the proper motion study by van Leeuwen et al. (2000).  Stars were given 
priority in the Hydra assignment program based on V magnitude, with a focus 
on stars in the range 11.0$<$V$<$14.0, which includes all giants in the cluster 
brighter than the HB up to the RGB tip.  Only stars with membership 
probabilities $\ga$80$\%$ were included for possible study.  All observations
took place between 2003 July 17 and 2003 July 19.  Three different Hydra setups
were used with exposure times ranging from 1800 to 3600 seconds.  Each setup 
allowed approximately 100 fibers to be placed on targets, yielding a total 
initial sample size of nearly 300 stars.  At V$\sim$13.5, reaching a 
signal--to--noise (S/N) ratio of 100 requires 3 hours of total integration 
time.  Unfortunately, weather and time constraints led to one of the setups 
receiving less than 2 hours of integration time with an average S/N of less 
than 50.  Many of these stars had to be excluded from analysis due to poor 
S/N; however, the final sample size still includes nearly 200 
stars.  These are shown in Figure \ref{f1} along with the complete sample 
given in van Leeuwen et al. (2000) for 11.0$<$V$<$14.0.

\begin{figure}
\epsscale{0.90}
\plotone{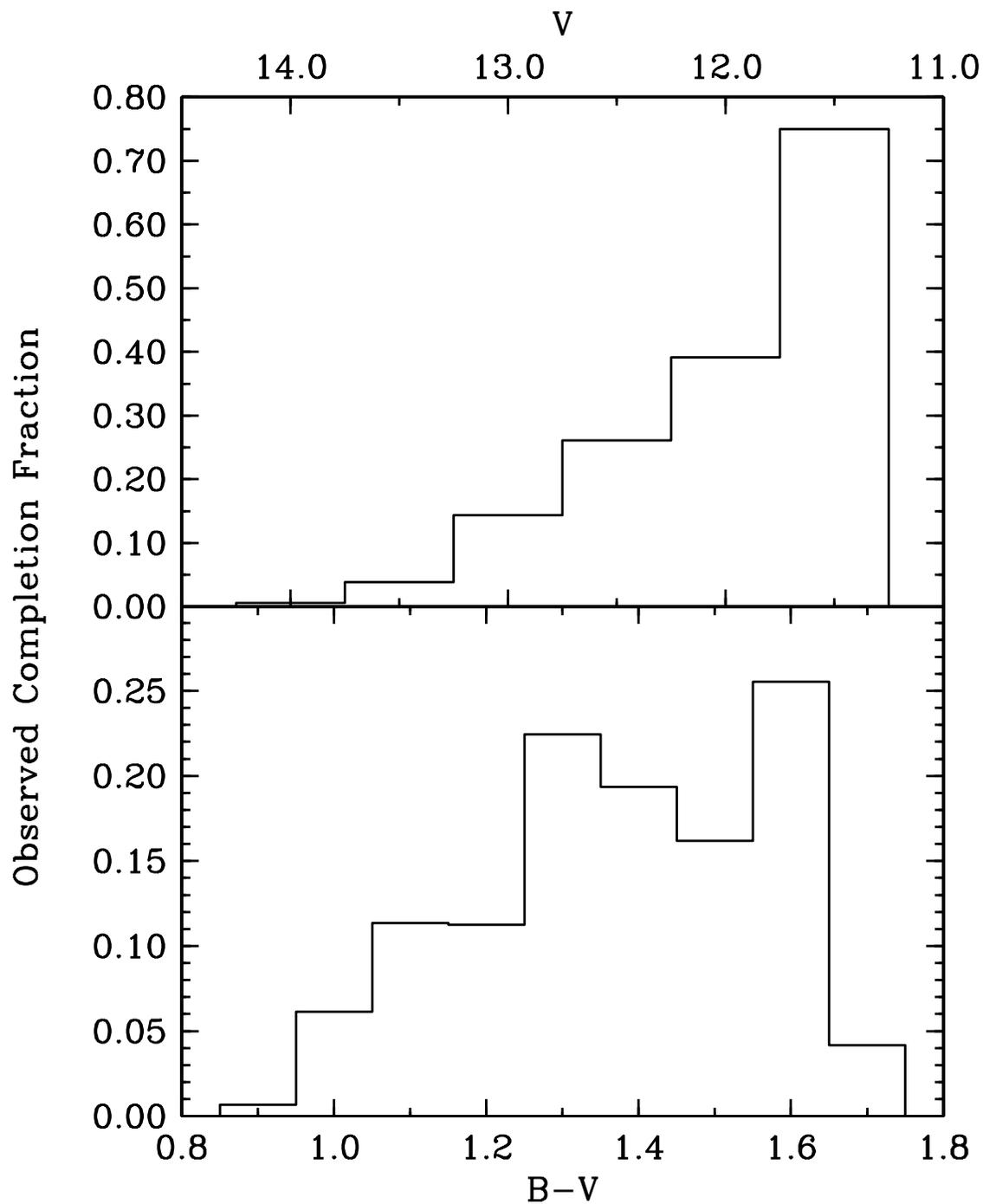}
\caption{Histogram showing the observed completion fraction of this study.  The
data are compared to the deeper photometric study of Rey et al. (2004).  The 
top panel shows the completion fraction binned by apparent V magnitude with
bin sizes of 0.5 mag. and the bottom panel shows the completion fraction binned
by B--V color in 0.1 mag. intervals.}
\label{f2}
\end{figure}

Due to $\omega$ Cen's broad RGB, selection effects must be taken into
account when interpreting abundance results.  Figure \ref{f2} shows our 
observed completion fraction of RGB stars both as a function of V magnitude
and B--V color compared to the deeper photometric study by Rey et al. 
(2004).  Since our observing program is biased towards 
selecting brighter stars, our sample includes more metal--poor than 
metal--rich stars because metal--rich stars have lower V magnitudes due to 
H$^{\rm -}$ opacity increasing with increasing metallicity.  While we observed 
75$\%$ of all RGB tip stars available, the fraction of stars observed 
decreases to $\sim$15--50$\%$ in the range 11.5$<$V$<$13.0.  Likewise, 
in considering completeness in B--V color, our sample includes stars of
higher luminosity for a given B--V, biasing our results towards the more 
metal--poor regime.

\begin{figure}
\epsscale{0.90}
\plotone{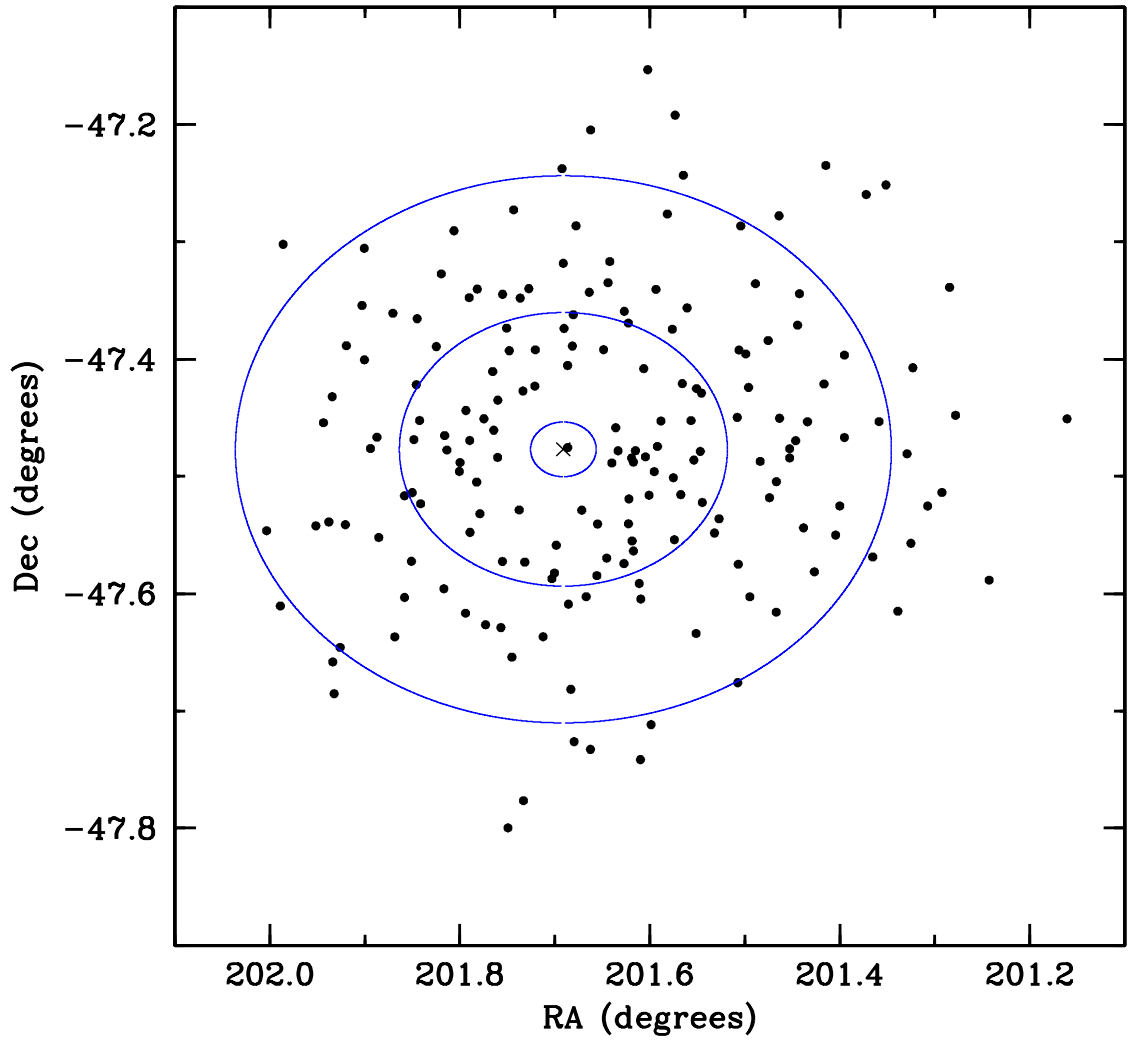}
\caption{Program stars are shown in terms of position in the field.  The cross 
indicates the field center at 201.691$\degr$, --47.4769$\degr$ (J2000)
(13$^{\rm h}$26$^{\rm m}$45.9$^{\rm s}$, --47$\degr$28$\arcmin$37.0$\arcsec$).
The ellipses indicate 1, 5, and 10 times the core radius of 1.40$\arcmin$.}
\label{f3}
\end{figure}

Figure \ref{f3} shows the location of our observed stars in right ascension and 
declination relative to the cluster center, defined by van Leeuwen et al. (2000)
as 13$^{\rm h}$26$^{\rm m}$45.9$^{\rm s}$, --47$\degr$28$\arcmin$37.0$\arcsec$
(J2000) and marked with a cross in the figure.  Since some evidence exists for 
a correlation between metallicity and distance from the cluster center 
(Norris et al. 1996; Suntzeff \& Kraft 1996; Norris et al. 1997; 
Hilker \& Richtler 2000; Pancino et al. 2000; Rey et al. 2004), we have 
observed stars as uniformly as possible at radii extending out to 
$\sim$20$\arcmin$.  Near the cluster center, crowding and the physical size of
the fibers limited the number of observations inside about 2 core radii,
where the core radius is approximately 1.40$\arcmin$ (Harris 1996; rev. 2003 
February).  We illustrate this effect with the ellipses in Figure \ref{f3} 
that correspond to 1, 5, and 10 core radii.

Basic data reductions were accomplished using the IRAF\footnote{IRAF is 
distributed by the National Optical Astronomy Observatories, which are 
operated by the Association of Universities for Research in Astronomy, Inc., 
under cooperative agreement with the National Science Foundation.} package 
\emph{ccdproc} to trim the bias overscan region and apply bias level 
corrections.  The IRAF task \emph{dohydra} was employed to correct for 
scattered light, extract the one--dimensional spectra, remove cosmic rays, 
apply a flat--field correction, linearize the wavelength scale, and subtract 
the sky spectrum.  Wavelength calibrations were carried out using a high S/N 
solar sky spectrum because the ThAr lamp was unavailable.  Standard IRAF tasks 
were used to co--add and normalize the spectra.  Typical S/N ratios for 
individual exposures ranged from $\sim$25--50, with co--added spectra having 
S/N between 75 and 150.

\section{RADIAL VELOCITY DETERMINATIONS}

$\omega$ Cen's location in the thick disk (Dinescu et al. 1999) makes field 
star contamination a more serious problem than for typical halo globular 
clusters.  While we initially only chose targets with high membership 
probabilities from van Leeuwen et al. (2000), direct measurements of target 
radial velocities assist with membership confirmation.  Radial velocities 
were determined using the IRAF tasks \emph{rvcor}, to correct for heliocentric 
motion, and \emph{fxcor}, to determine the heliocentric radial velocity.  For 
the comparison spectrum, we used the same high S/N daylight sky spectrum that 
was used for wavelength calibration.  A summary of our determined radial 
velocities along with membership probabilities from van Leeuwen et al. (2000)
are given in Table 2.

The largest radial velocity study of $\omega$ Cen stars to date is by Reijns
et al. (2006), who determined radial velocities for $\sim$2,000 RGB stars.
Their study finds a strongly peaked distribution near 232 km s$^{\rm -1}$, with
a median uncertainty of less than 2 km s$^{\rm -1}$ and a velocity dispersion
of $\sim$6 km s$^{\rm -1}$ for the inner 25$\arcmin$ of the cluster.  Similarly,
Mayor et al. (1997) find $\langle$V$_{\rm R}$$\rangle$=232.8 $\pm$ 0.7 km
s$^{\rm -1}$ ($\sigma$$\sim$17.5 km s$^{\rm -1}$) for 471 stars and 
Suntzeff \& Kraft (1996) find $\langle$V$_{\rm R}$$\rangle$=234.7 $\pm$ 1.3 
km s$^{\rm -1}$ ($\sigma$=11.3 km s$^{\rm -1}$) for their ``bright" sample of 
199 stars.  Recently, Pancino et al. (2007) determined radial velocities for
650 RGB stars and found $\langle$V$_{\rm R}$$\rangle$=233.4 $\pm$ 0.5 km
s$^{\rm -1}$ ($\sigma$=13.2 km s$^{\rm -1}$).  We find in agreement with 
these studies: $\langle$V$_{\rm R}$$\rangle$=231.8 km s$^{\rm -1}$ $\pm$ 1.6 
km s$^{\rm -1}$ ($\sigma$=11.6 km s$^{\rm -1}$).  Our observations do not 
provide an absolute velocity calibration, but comparison with the other 
observations of the average velocity of cluster stars suggests that the 
systematic error of our radial velocities is less than about 2 km s$^{\rm -1}$.
Since all of our stars listed in Table 2 are less than 3$\sigma$ away from the 
cluster averaged velocity and $\omega$ Cen's velocity is high relative to the 
general field population, it is unlikely any of our targets are field stars.

\section{Analysis}

We have derived Fe and Al abundances using lines available in the spectral 
range 6500--6750~\AA\ with either equivalent width or synthetic spectrum
analyses.  Spectrum synthesis was used to determine Al abundances in 
metal--rich and/or CN--strong stars.  When multiple lines were available, the
stated abundances represent the average of the individual lines.  Effective 
temperatures (\emph{T$_{\rm eff}$}) and gravities (\emph{log g}) were 
estimated using published (V--K)$_{\rm 0}$ photometry.  \emph{T$_{\rm eff}$} 
and microturbulence (\emph{V$_{\rm t}$}) were further refined via 
spectroscopic analyses.  The analysis follows the methods described in Johnson 
et al. (2005) and Johnson \& Pilachowski (2006).

\subsection{Model Stellar Atmospheres}

Using V photometry from van Leeuwen et al. (2000) and K$_{\rm s}$ photometry 
from 2MASS, we estimated \emph{T$_{\rm eff}$} with the color--temperature 
relation described in Alonso et al. (1999; 2001), which is based on the 
infrared flux method (Blackwell \& Shallis 1977).  However, the Alonso et al. 
(1999) method requires the photometry to be on the Carlos S\'{a}nchez 
Telescope (TCS) photometric system.  We transformed the V and K$_{\rm s}$ 
magnitudes onto the TCS system using the transformations provided in Alonso et 
al. (1994; 1998) and Carpenter (2001), as summarized in Johnson et al. (2005).
To correct for interstellar reddening and extinction, we applied the 
correction recommended by Harris (1996; rev. 2003 February) of E(B--V)=0.12 
and used E(V--K)/E(B--V)=2.7 (Johnson 1965).  While Calamida et al. (2005) claim
differential reddening, perhaps differing by as much as a factor of two near 
the core, could be a problem, the well defined evolutionary sequences seen in 
Villanova et al. (2007) seem to indicate it is not too severe.  Van Loon et al.
(2007) find some evidence for interstellar absorption by gas in the cluster, but
this is concentrated near the core where our observations are sparse.  
Therefore, we have only applied a uniform reddening correction.  Bolometric 
corrections were applied using the empirical relations given in Alonso et al. 
(1999) assuming a distance modulus of (m--M)$_{\rm V}$=13.7 (van de Ven et al. 
2006).

Applying the proper color--temperature relation requires knowledge of a star's 
metallicity.  Therefore, we took the empirical relation given in van Leeuwen 
et al. (2000; their eq. 15), which gives [Ca/H] as a function of V and B--V, 
and assumed [Ca/Fe]$\sim$+0.30 for [Fe/H]$\la$--1.0 (e.g., Norris \& Da Costa 
1995), with a linear decrease towards [Ca/Fe]=0.0 at [Fe/H]=0.0.  This gave a 
rough estimate of [Fe/H] for each star and allowed us to choose the proper 
equation in Alonso et al. (1999).  

Since only one Fe II line was available for analysis (6516~\AA), we determined
surface gravity using the standard relation,
\begin{equation}
log(g)=0.40(M_{bol.}-M_{bol.\sun})+log(g_{\sun})+4(log(T/T_{\sun}))+
log(M/M_{\sun}),
\end{equation}
instead of the ionization equilibrium of Fe.  We assumed M=0.80 M$_{\sun}$ for
all stars, regardless of metallicity.  Though there may be an intrinsic age 
spread of a few Gyr on the RGB (see $\S$5 for further discussion on this 
issue), this will lead to a mass difference only of order 
$\sim$0.05 M$_{\sun}$, which is negligible for surface gravity determinations.

In addition to \emph{T$_{\rm eff}$}, \emph{log g}, and [Fe/H] estimates, we 
also needed a starting point with \emph{V$_{\rm t}$}.  Initial estimates were 
based on the empirical relation derived in Pilachowski et al. (1996), which 
gives \emph{V$_{\rm t}$} as a function of \emph{T$_{\rm eff}$} for metal--poor 
field giants and subgiants.  Typical \emph{V$_{\rm t}$} values ranged from 
about 1.3--2.3 km s$^{\rm -1}$ in the temperature range 5000--3800 K, 
respectively.

\begin{figure}
\epsscale{0.90}
\plotone{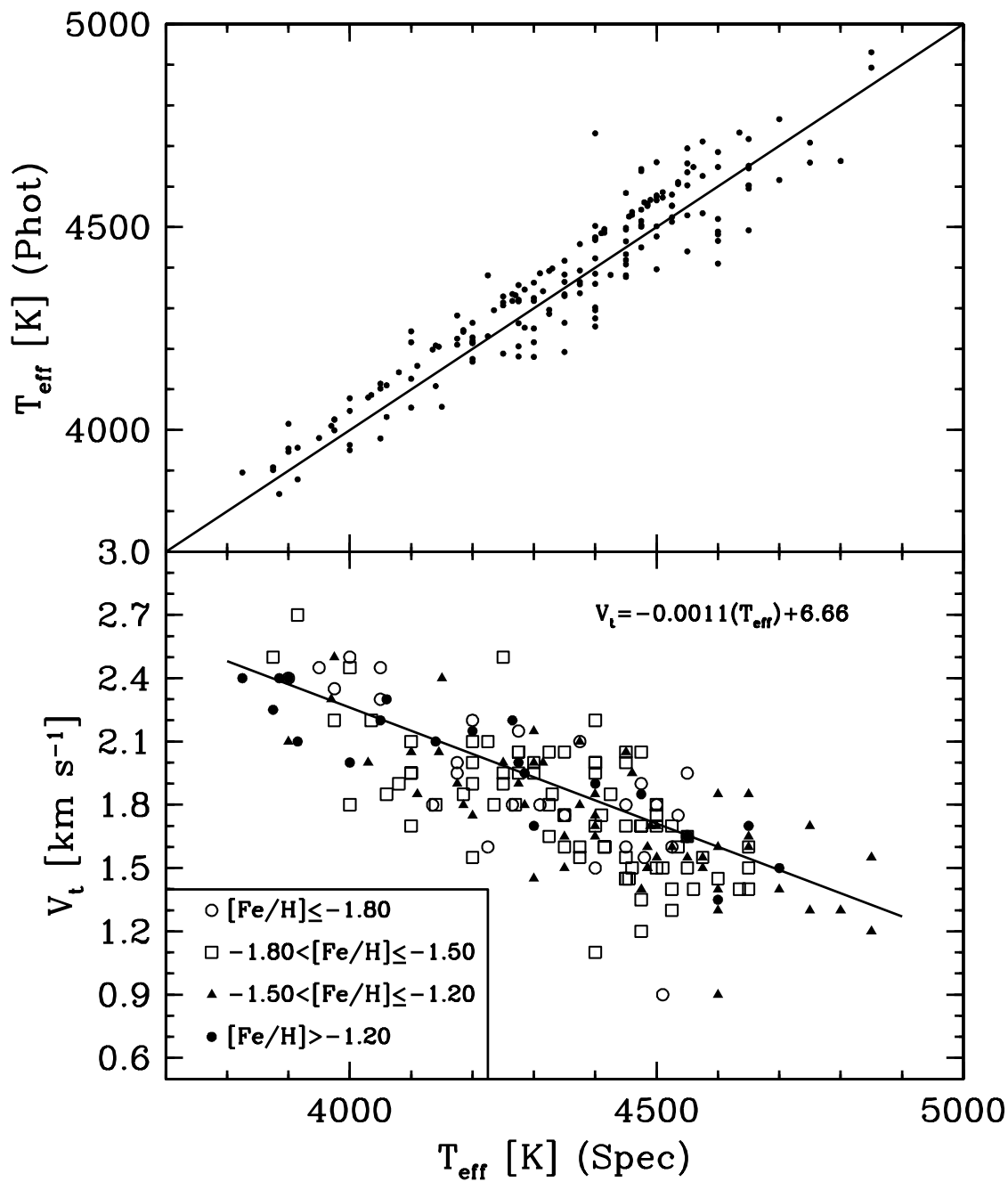}
\caption{The top panel shows the relation between the effective temperature 
estimated via V--K photometry versus the spectroscopically determined 
temperature.  The straight line indicates perfect agreement.  The bottom panel
illustrates microturbulent velocity versus effective temperature.  Different
symbols indicate stars in different metallicity bins as indicated above.  A 
linear least--squares fit is provided along with the equation relating 
microturbulence to effective temperature.}
\label{f4}
\end{figure}

\begin{figure}
\epsscale{0.90}
\plotone{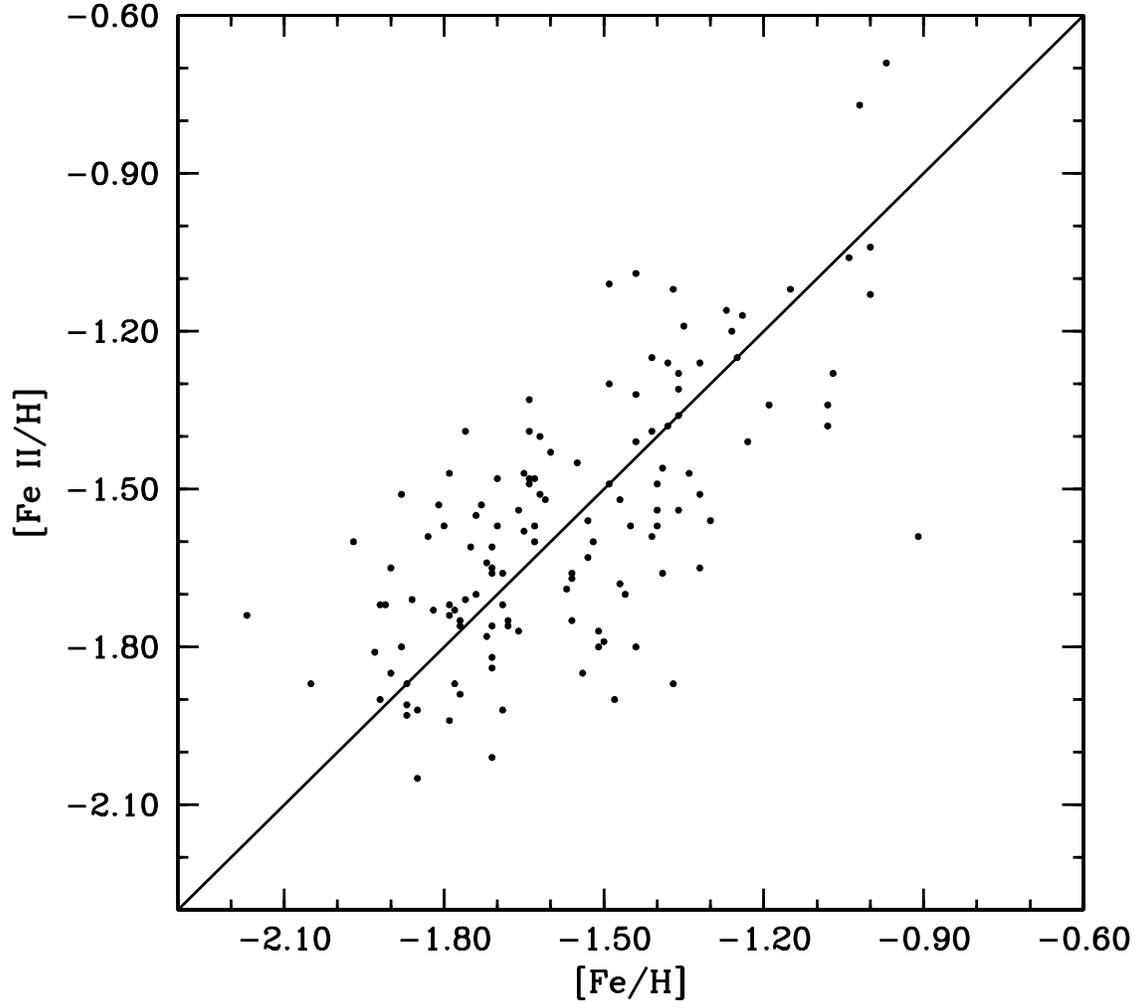}
\caption{Derived [Fe II/H] abundances are plotted versus [Fe I/H].  The line 
indicates perfect agreement.}
\label{f5}
\end{figure}

We generated the model stellar atmospheres by interpolating in the 
ATLAS9\footnote{The model atmosphere grids can be downloaded from 
http://cfaku5.cfa.harvard.edu/grids.html.} (Castelli et al. 1997) grid of 
models without convective overshoot.  Initial models were created using the 
\emph{T$_{\rm eff}$}, \emph{log g}, [Fe/H], and \emph{V$_{\rm t}$} estimates 
as described above.  \emph{T$_{\rm eff}$} was further refined by removing 
trends in Fe abundance as a function of excitation potential.  Likewise, 
\emph{V$_{\rm t}$} was improved by removing trends in Fe abundance as a 
function of reduced width (log(EW/$\lambda$)).  A comparison between 
photometric and spectroscopically determined temperatures is given in the top 
panel of Figure \ref{f4}.  Typical photometric and spectroscopic temperature 
estimates agree to within approximately $\pm$100 K.  The bottom panel of 
Figure \ref{f4} shows our spectroscopically determined \emph{V$_{\rm t}$} as a 
function of \emph{T$_{\rm eff}$} for different metallicity bins with a linear 
least squares fit given by,
\begin{equation}
V_{\rm t}=-0.0011(T_{\rm eff})+6.66,
\end{equation}
which is independent of metallicity.  This fit agrees to within 
$\sim$0.10--0.15 km s$^{\rm -1}$ to that given in Pilachowski et al. (1996).  
Figure \ref{f5} shows our derived [Fe II/H] given as a function of [Fe/H].
As stated above, we only had one Fe II line available for analysis, but
the fact that both Fe estimates agree to within 0.16 dex on average 
($\sigma$=0.12 dex) leads us to believe our surface gravity estimates are not in
serious error.  A complete list of our adopted model atmosphere parameters is
provided in Table 3.

\subsection{Derivation of Abundances}

Abundances were determined using equivalent width analyses for all Fe lines 
and most Al lines, with the exception of cases where evidence for considerable
CN contamination near the 6696, 6698~\AA\ Al doublet (i.e., metal--rich and/or
CN--strong stars) existed and spectrum synthesis was used instead.  We 
measured equivalent widths using a FORTRAN program developed for this project 
that interactively fits a Gaussian curve to each absorption line by 
implementing a Levenberg-Marquardt algorithm (Press et al. 1992) to find the 
least--squares fit given a continuum level and limits of integration.  A 
high resolution, high S/N Arcturus spectrum\footnote{The Arcturus Atlas can be
downloaded from the NOAO Digital Library at 
http://www.noao.edu/dpp/library.html.} was simultaneously overplotted for each 
spectrum to aide in continuum placement and line identification.  The program 
also has the ability to fit up to five Gaussians simultaneously for deblending 
purposes; however, all equivalent widths were verified independently using 
IRAF's \emph{splot} package.  

\subsubsection{Equivalent Width Analysis}

Final abundances were calculated using the abfind driver in the 2002 version 
of the local thermodynamic equilibrium line analysis code MOOG (Sneden 
1973).  Adopted log gf values are the same as those employed in Johnson et al. 
(2006), which were adapted from line lists provided in Th{\'e}venin (1990), 
Sneden et al. (2004; modified from Ivans et al. 2001), and Cohen \& 
Mel{\'e}ndez (2005).  A summary of our line list is given in Table 4 and the 
measured equivalent widths are provided in Table 5.

While we had identified 20 Fe I lines for analysis, in most cases only 10--15
lines could be used due to severe line blending, bad ccd pixels, or line 
strength.  In this sense, only lines lying on the linear part of the curve
of growth were used, which meant neglecting almost all lines with a reduced 
width larger than about --4.5 (roughly 200 m\AA\ at 6600~\AA).  This
unfortunately meant that many lines in metal--rich stars are too strong to give 
accurate abundances using our techniques.  For the cases where Al abundances 
were determined using equivalent width measurements, weak line blends were 
taken into account using deblending methods.  As stated above, stars with 
strong line blending or molecular line blanketing in the region near the Al 
doublet were analyzed with spectrum synthesis.

\begin{figure}
\epsscale{0.90}
\plotone{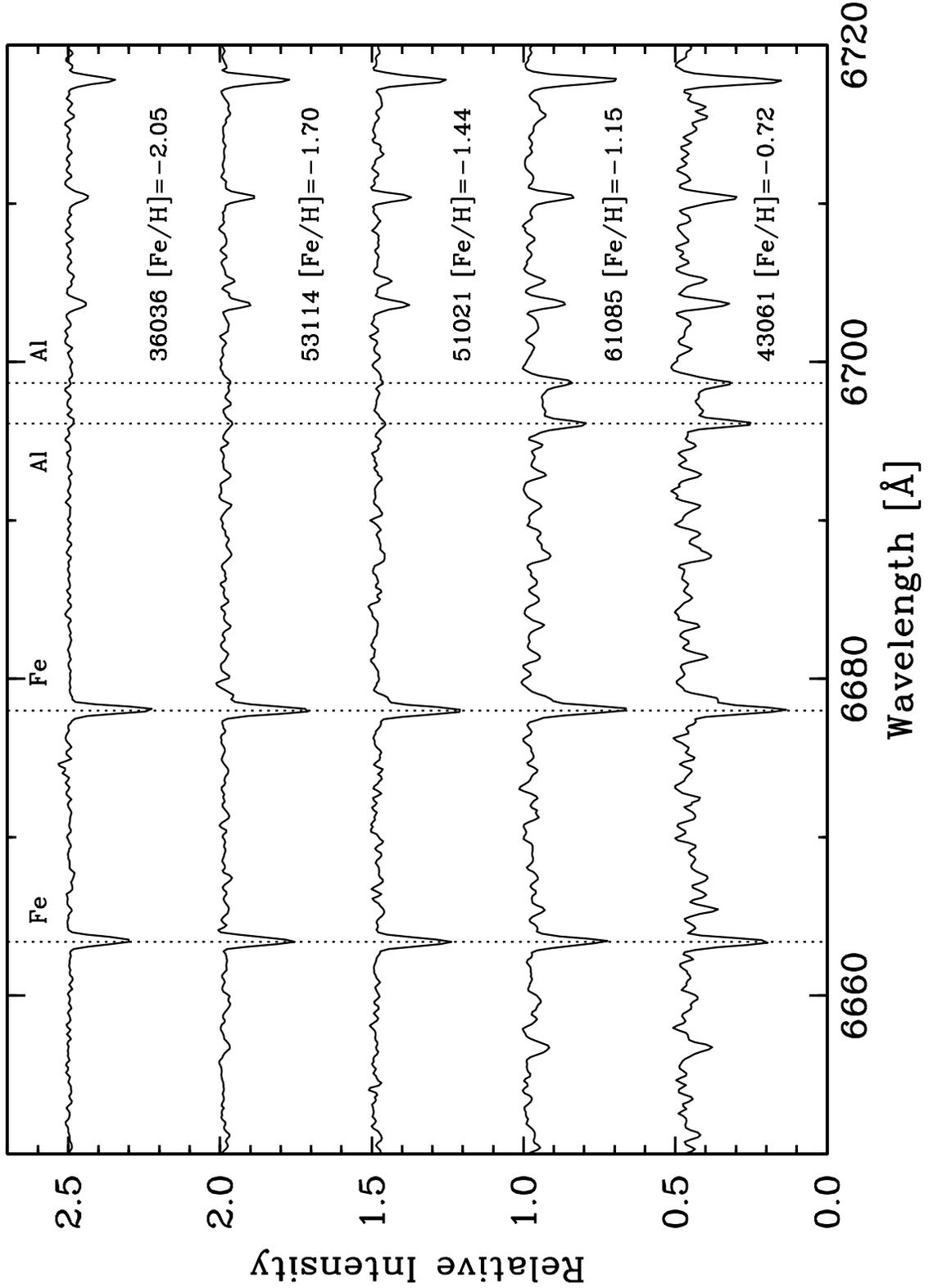}
\caption{Several sample spectra are shown for various [Fe/H].  The spectra have
been offset for display purposes.  For reference the vertical dashed lines 
indicate the location of the Al I lines and two additional Fe I lines.  From
top to bottom, the [Al/Fe] values for the stars shown are +0.40, +0.45, +0.15,
+0.97, and +0.57, respectively.}
\label{f6}
\end{figure}

Typical uncertainties are quite small for [Fe/H] determinations with internal 
line--to--line spreads of $\sim$0.10--0.15 dex and $\sigma$/$\sqrt{N}$ $<$ 0.05
dex on average.  Sample spectra for stars of approximately the same 
\emph{T$_{\rm eff}$} but different metallicities are shown in Figure \ref{f6}.
Here we illustrate that our [Fe/H] determinations are at the very least 
consistent in a relative sense as one notices the increasing Fe line strengths 
and CN--band strengths with increasing metallicity.  The uncertainty in Al 
abundances is larger given that only two lines are available, but the two 
lines give a remarkably consistent abundance, with an average 
$\sigma$/$\sqrt{N}$=0.08 dex.  It should be noted that in several of our 
spectra only one Al line could be confidently measured due mostly to bad 
pixels.  In Figure \ref{f6}, the reader can see the stark contrast in line 
strength between a star such as 51021, which has [Al/Fe]=+0.15 at 
[Fe/H]=--1.44, and star 61085, which has [Al/Fe]=+0.97 at [Fe/H]=--1.15.  A 
summary of all derived abundances and associated $\sigma$/$\sqrt{N}$ values is 
given in Table 6.

\subsubsection{Spectrum Synthesis Analysis}

As mentioned above, we determined Al abundances for metal--rich and/or 
CN--strong stars using the synth driver in MOOG.  Candidates for spectrum
synthesis were chosen based on visual inspection of the 6680--6700~\AA\
region, where the majority of lines surrounding the Al doublet are CN lines.
Stars where CN contamination was seen between the Al lines were designated
for synthetic spectrum analysis (e.g., see Figure 6, lower two spectra).

The atomic line list (with the exception of the two Al lines) was taken from the
Kurucz atomic line database\footnote{The Kurucz line list database can be 
accessed via http://kurucz.harvard.edu/linelists.html.}.  We adjusted the 
oscillator strengths from this line list so the line strengths matched those 
in the solar spectrum.  For the CN molecular line list, we used a combination 
of one available from Kurucz and one provided by Bertrand Plez (2007, private 
communication; for a description on how the line list was prepared, see Hill
et al. 2002).  

Since most of the program stars do not have known C, N, or 
\iso{12}{C}/\iso{13}{C} abundances, we started with [C/Fe]=--0.5, [N/Fe]=+1.5,
and \iso{12}{C}/\iso{13}{C}=5, values roughly consistent with previous work
(e.g., Norris \& Da Costa 1995; Smith et al. 2002).  We then treated the 
nitrogen abundance as a free parameter and adjusted it until a satisfactory 
fit was achieved.  Typical best fit [N/Fe] values were $\sim$+1.0 to +1.5.  
To test the effect of different \iso{12}{C}/\iso{13}{C} ratios, we generated 
two sets of spectra with \iso{12}{C}/\iso{13}{C}=5 and 
\iso{12}{C}/\iso{13}{C}=1000.  The fits to the CN lines were indistinguishable 
between the two cases, meaning \iso{12}{C} is the dominant isotope in this 
spectral region and thus synthesized CN lines are insensitive to the 
\iso{13}{C} abundance.

\begin{figure}
\epsscale{0.90}
\plotone{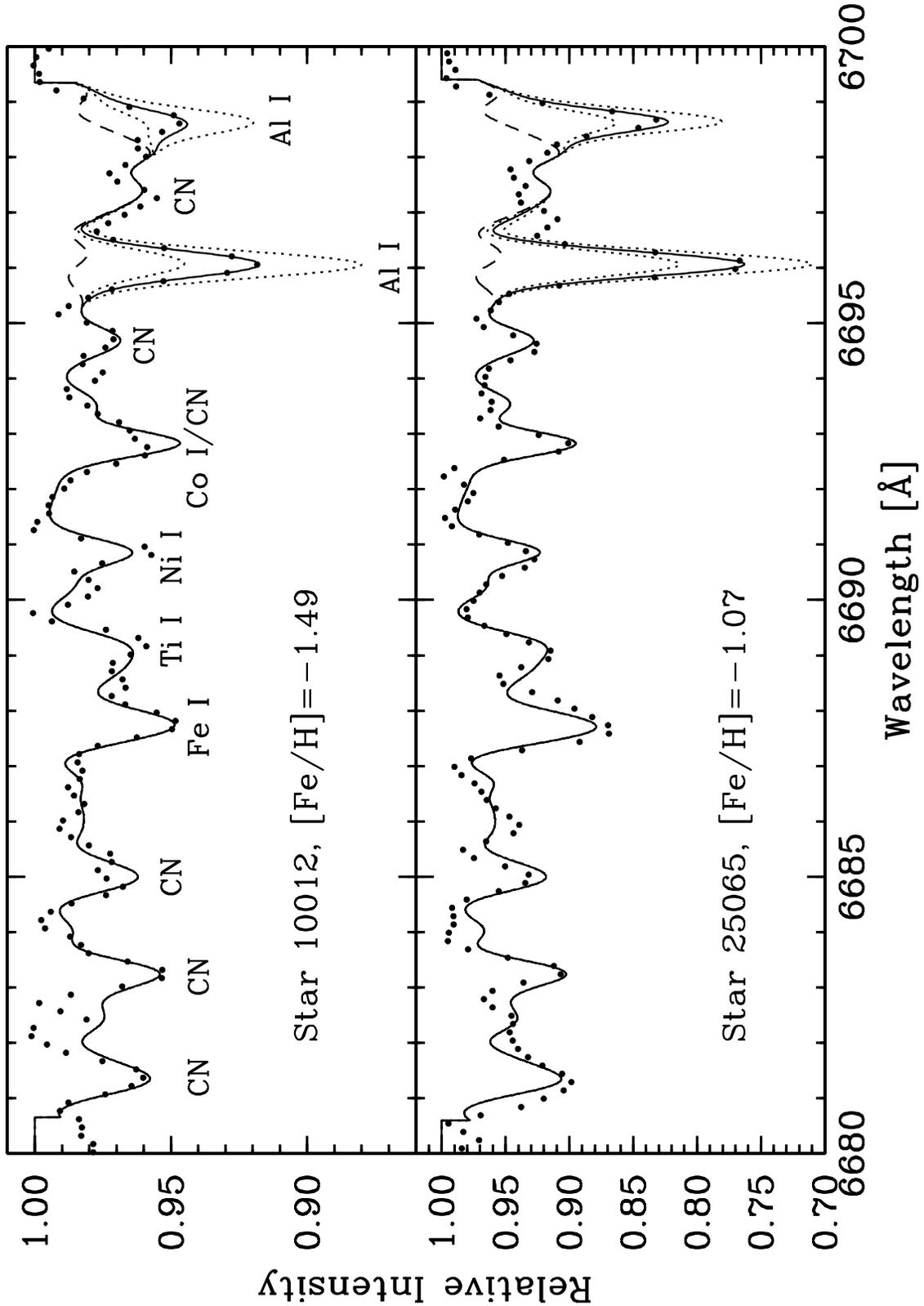}
\caption{Sample spectrum syntheses of the Al region are shown.  The dashed 
line indicates log $\epsilon$(Al)=--5.0, the solid line shows the best--fit Al
abundance, and the dotted lines indicate abundance $\pm$0.30 dex from the 
best--fit Al value.}
\label{f7}
\end{figure}

With the CN lines fit, we were then able to adjust the Al abundance until 
the synthetic spectrum matched the observed.  Sample synthesis fits are given
in Figure \ref{f7} for a metal--poor and metal--rich case.  Aside from the CN
lines, the Fe I line near the 6696~\AA\ feature is the only other contaminating
line in the region, but this line has an excitation potential of nearly
5 eV, making its contribution mostly negligible in these cool stars.  
Generally, the abundances given by the 6696 and 6698~\AA\ lines agreed to 
within about $\pm$0.10 dex.  Since a significant percentage of our Al 
abundances were determined using synthesis analyses, we tested for systematic
offsets between synthesis and equivalent width methods.  For sample stars that
were both metal--poor and did not show signs of CN contamination, the 
difference in [Al/Fe] determined via both methods was less than 0.05 dex.  
However, for higher metallicity stars and those with possible CN contamination,
the difference was 0.10--0.20 dex, with equivalent width analyses always 
overestimating the abundance.  The quoted values for Al abundances derived via 
spectrum synthesis are given as the average from those two lines.  A summary of 
our derived abundances is given in Table 6.  Stars with Al determinations 
via synthesis are designated by ``Syn" in the 6696 and 6698 \AA\ columns of
Table 5.

\subsubsection{Abundance Sensitivity to Model Atmosphere Parameters}

We tested the effects on derived abundances from changes in model atmosphere
parameters by altering \emph{T$_{\rm eff}$} $\pm$ 100 K, \emph{log g} $\pm$ 
0.25 cm s$^{\rm -2}$, and \emph{V$_{\rm t}$} $\pm$ 0.25 km s$^{\rm -1}$ for 
models of [Fe/H]=--2.0, --1.5, and --1.0.  As can be seen in Table 7, 
\emph{T$_{\rm eff}$} uncertainties are the primary source of error for Fe I 
and Al I, and surface gravity is the primary source for Fe II abundances.  
This seems logical given that Fe I and Al I reside in a subordinate ionization 
state, and Fe II exists in the primary ionization state.  

Following Table 7, an uncertainty of order 100 K in \emph{T$_{\rm eff}$} leads 
to an error of $\sim$0.10--0.20 dex in Fe I, though the effect is somewhat 
reduced at higher metallicity.  The opposite is true for Fe estimates based
solely on the Fe II line, where the error range is $\sim$0.05--0.10 dex and 
the uncertainty becomes larger with increasing metallicity.  Though the 
variation in Al I abundance as a function of \emph{T$_{\rm eff}$} uncertainty 
is smaller than for Fe I, it is still of order 0.10 dex with a weak dependence
on metallicity.

The effects of surface gravity uncertainty are of order 0.10 dex for the Fe II 
line, but are negligible for the neutral Fe and Al lines.  For this reason,
enforcing ionization equilibrium between different species is often used for
constraining surface gravity estimates.  As mentioned in $\S$4.2.1, having only
one Fe II line means the Fe abundance derived from Fe II is probably no more
accurate than the typical line--to--line scatter present in Fe I 
($\sigma$$\sim$0.10--0.15 dex).  Combined with the sensitivity of Fe II to 
surface gravity estimates of order $\pm$0.25 cm s$^{\rm -2}$, the fact that 
agreement between Fe I and Fe II is better than about 0.10 dex (see 
Figure \ref{f5}) suggests estimates based on evolutionary arguments provide a 
decent approximation to the surface gravity; however, Table 7 shows this has 
little effect on our derived Fe I and Al I abundances.  From this, we can 
safely assume that contamination from AGB stars, which have M$\sim$0.60 
M$_{\sun}$ and thus a lower surface gravity, will not significantly alter our 
results.

The ad hoc microturbulence parameter, adjusted to remove abundance trends as 
a function of reduced width, has the strongest effect for lines lying on the
flat part of the curve of growth.  As is seen in Table 7, the effect on the Fe I
abundance due to uncertainty in \emph{V$_{\rm t}$} increases with increasing 
metallicity because the lines become progressively stronger.  However, Fe II 
and Al I are mostly unaffected due to their relatively small equivalent widths 
and the effect on Fe I is still $<$0.10 dex even at [Fe/H]=--1.0.

In addition to variations in model stellar atmosphere parameters we tested the 
sensitivity of Al abundance to CN strength via spectrum synthesis by varying 
[N/Fe]$\pm$0.30 dex.  Changing the nitrogen abundance by this amount worsens 
the fit to the CN lines in the spectrum, but alters the derived [Al/Fe] 
abundance less than 0.10 dex at all metallicities.  Note that since [O/Fe] is 
unknown for most of our program stars and [O/Fe] can have values ranging from 
about +0.30 to less than --0.50, it is not possible to constrain the molecular
equilibrium equations to derive true [C/Fe] and [N/Fe].  We present the 
[Al/Fe] results for each metallicity bin in Table 7.

\subsection{Comparison with the Literature}

While $\omega$ Cen has been the subject of multiple abundance studies 
(see $\S$ 1 for a brief review), most of these are low resolution studies that 
do not involve elements other than Fe and/or Ca.  Therefore, we are only 
comparing results in the literature for which moderate to high resolution 
Al data are available and with which we have three or more stars in common.
This limits the comparison to Brown \& Wallerstein (1993; 3 stars), Norris \& 
Da Costa (1995; 24 stars), Zucker et al. (1996; 4 stars), and Smith et al.
(2000; 3 stars).

\begin{figure}
\epsscale{1.00}
\plotone{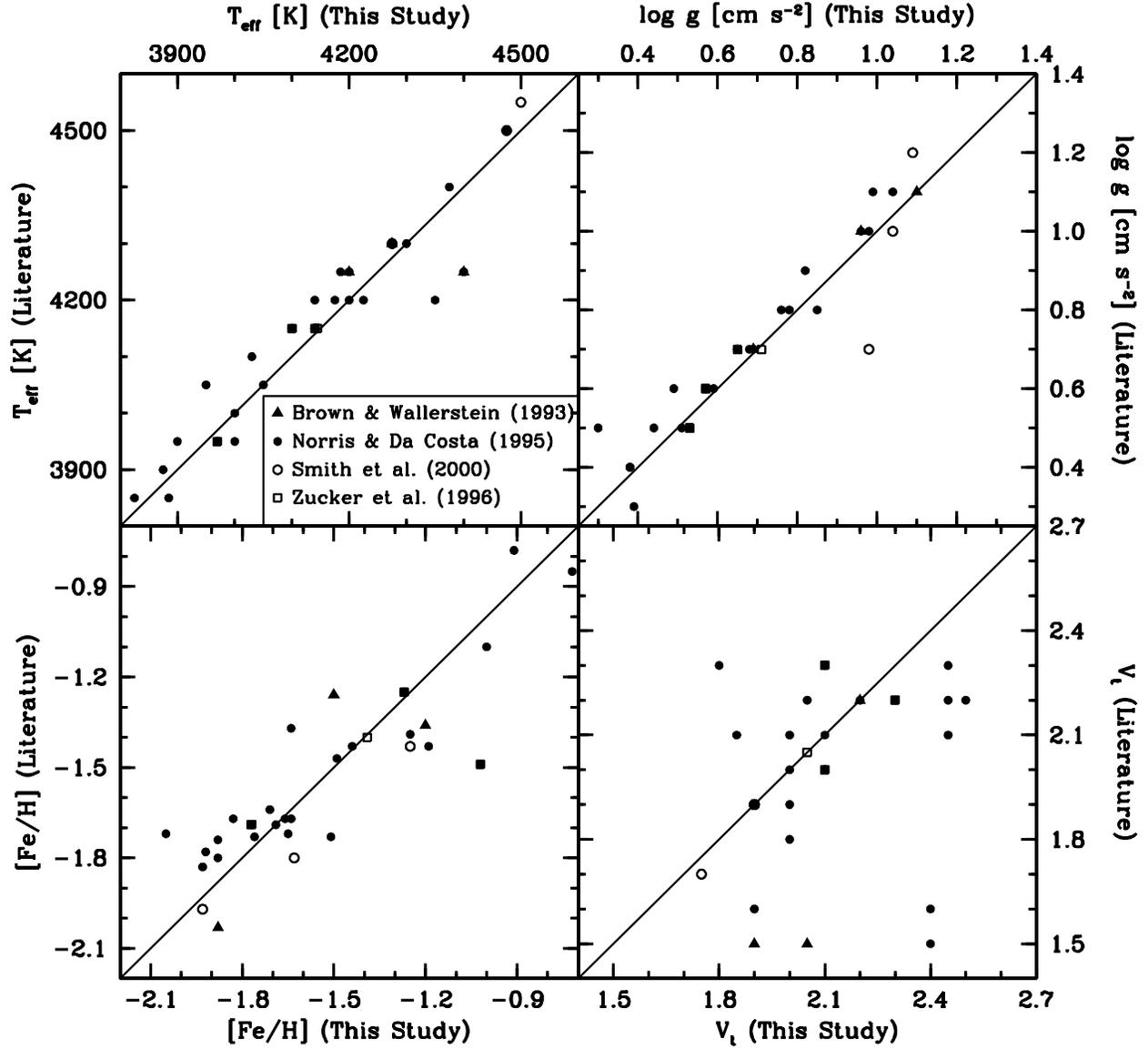}
\caption{The four panels show our adopted model atmosphere parameters versus 
those available in the literature.  A straight line indicates perfect agreement
in all panels.}
\label{f8}
\end{figure}

In Figure \ref{f8}, we present the values of \emph{T$_{\rm eff}$}, 
\emph{log g}, [Fe/H], and \emph{V$_{\rm t}$} given in the literature versus 
those obtained in this study.  As can be seen from the figure, agreement is 
quite good for the temperature and surface gravity estimates, with the scatter 
increasing slightly for the metallicity and microturbulence estimates.  For 
\emph{T$_{\rm eff}$}, the average offset between our study and the literature 
is --7 K ($\sigma$$\sim$50 K), and the average difference for \emph{log g} 
is --0.02 cm s$^{\rm -2}$ ($\sigma$$\sim$0.10 cm s$^{\rm -2}$).  This 
indicates that any disagreement between literature Fe and Al abundances and 
ours is not due to choices of \emph{T$_{\rm eff}$} and \emph{log g}.  
Similarly, [Fe/H] measurements agree to within 0.02 dex on average 
($\sigma$$\sim$0.20 dex).  The reason for the larger dispersion in 
microturbulence estimates is not entirely clear, but it could be due to factors
such as the number of lines available, data quality, continuum placement, and 
type of lines used (i.e., high and/or low excitation potential).  However,
on average the agreement is within 0.10 km s$^{\rm -1}$ ($\sigma$$\sim$0.25 km
s$^{\rm -1}$).

\begin{figure}
\epsscale{1.00}
\plotone{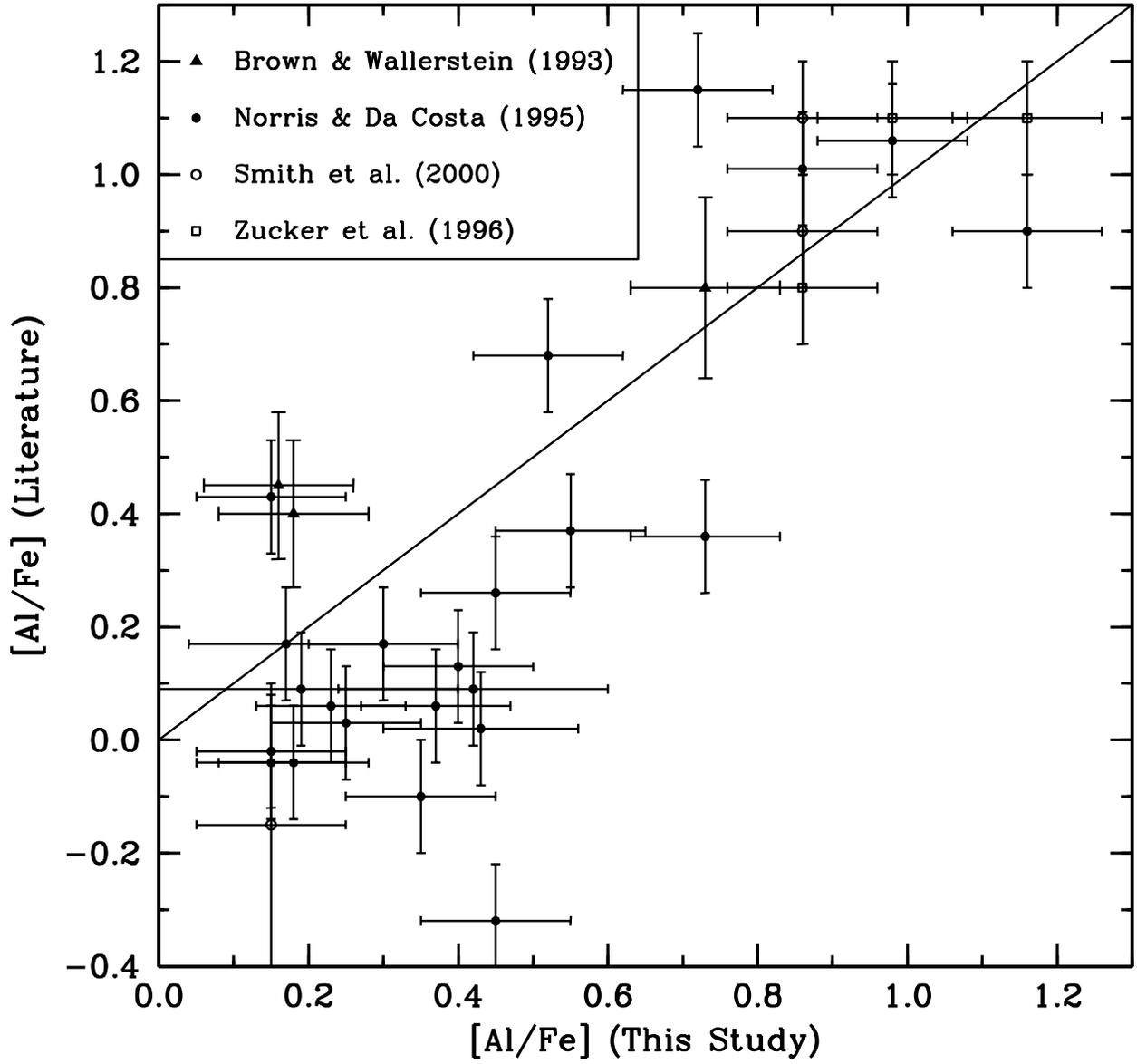}
\caption{Al abundances available in the literature are plotted versus those 
derived here.  The straight line indicates perfect agreement.  The error bars
are those given from each study and this one.  If no error is provided, a base
value of $\pm$0.10 dex is assumed.}
\label{f9}
\end{figure}

Comparison between our derived [Al/Fe] abundances versus those in the 
literature are provided in Figure \ref{f9}.  Given the various data qualities,
choices of model atmospheres and parameters, and adopted atomic line data, 
agreement is again quite good.  The average offset between our derived 
abundances and those available in the literature is 0.06 dex 
($\sigma$$\sim$0.30 dex).  Given that typical uncertainties in [Al/Fe] are of 
order 0.10--0.20 dex, agreement is comparable to that range. 

\section{RESULTS AND DISCUSSION}

\subsection{Fe Abundances}

As discussed in $\S$1, it has been known for many years and shown by several 
authors that $\omega$ Cen has a considerable spread in metallicity that ranges
from slightly less than [Fe/H]=--2.0 to more than [Fe/H]=--0.7.  While several
lower resolution spectroscopic (Norris et al. 1996; Suntzeff \& Kraft 1996; 
Sollima et al. 2005b; Kayser et al. 2006; Stanford et al. 2006; Stanford et al.
2007; van Loon et al. 2007\footnote{The referee noted discrepancies between 
the [Fe/H] values derived by Norris \& Da Costa (1995) and van Loon et al. 
(2007).  We note that our results agree with Norris \& Da Costa and a detailed 
resolution of this problem is beyond the scope of this paper.}; Villanova et 
al. 2007) and photometric (Lee et al. 1999; Hilker \& Richtler 2000; Hughes \& 
Wallerstein 2000; Pancino et al. 2000; van Leeuwen et al. 2000; Rey et al. 
2004; Stanford et al. 2004; Sollima et al. 2005a; Stanford et al. 2006) 
studies have obtained metallicity estimates for a large number of stars 
(N$\ga$500 in some cases), there have only been a few high resolution 
spectroscopic studies with a significant number (N$\ga$10) of stars observed 
(Norris \& Da Costa 1995; Smith et al. 2000; Piotto et al. 2005; Sollima et 
al. 2006).  However, aside from the present study, Norris \& Da Costa (1995) 
still represents the largest (N=40) single high resolution analysis of 
$\omega$ Cen RGB stars.  The general results from the metallicity studies can 
be summarized as: (1) few stars exist at [Fe/H]$<$--2.0, (2) a primary peak in 
the metallicity distribution is observed at [Fe/H]$\sim$--1.8 to --1.6, (3) 
there is a long tail of increasing metallicity up to [Fe/H]$\sim$--0.5, and 
(4) there appear to be multiple peaks in the distribution at various [Fe/H] 
values.

\begin{figure}
\epsscale{1.00}
\plotone{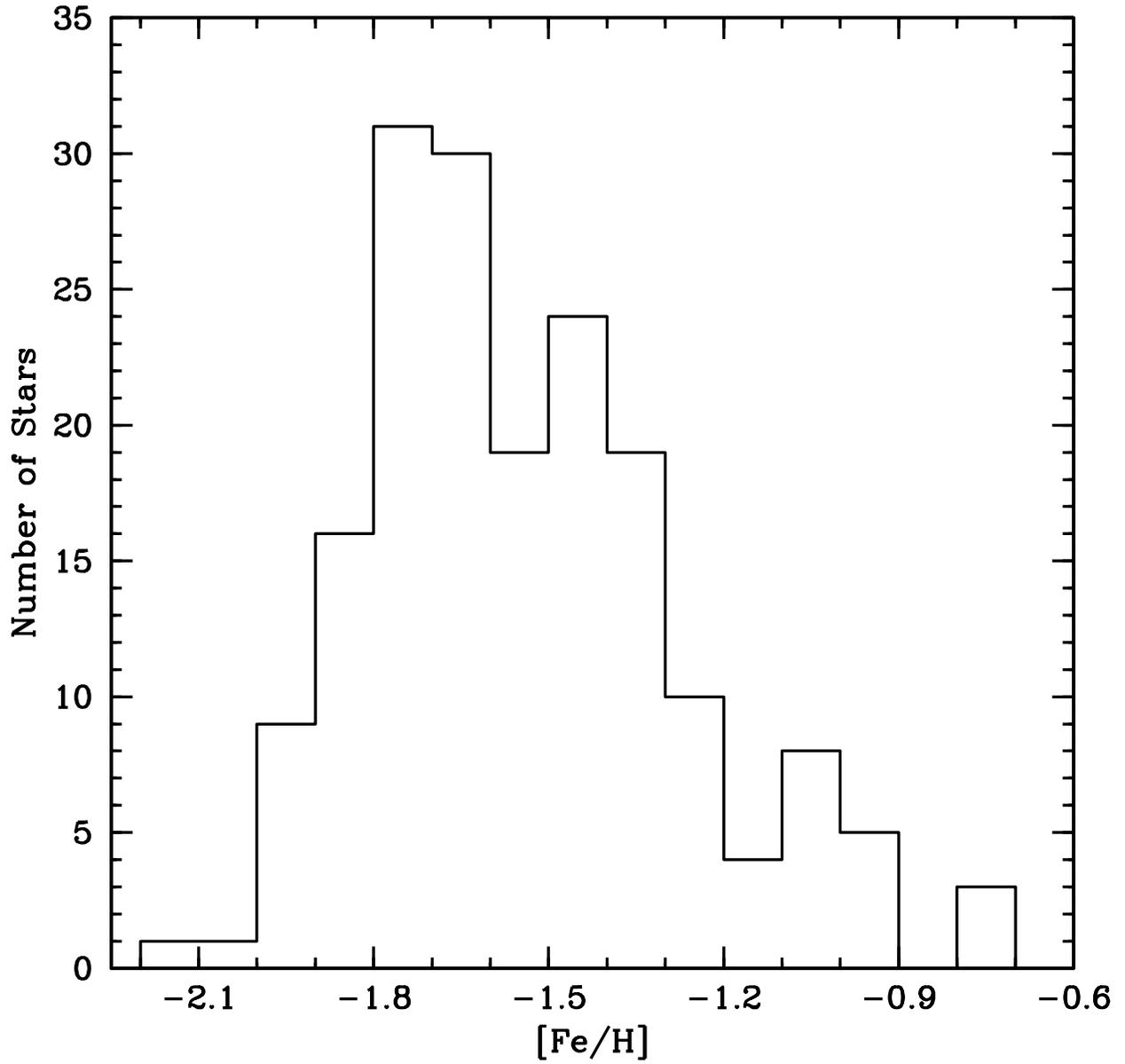}
\caption{A histogram of derived [Fe/H] values with bin sizes of 0.10 dex.}
\label{f10}
\end{figure}

In Figure \ref{f10}, we present a histogram of our derived metallicity 
distribution function for all 180 stars.  We find in agreement with previous 
studies that there are at least four distinct populations with the most 
metal--poor having [Fe/H]$\sim$--1.75, the two intermediate metallicity
populations have [Fe/H]$\sim$--1.45 and --1.05, and the most metal--rich 
population has [Fe/H]$\sim$--0.75.  While our observations are skewed towards
observing more metal--poor stars (see Figure \ref{f2}), there are
intrinsically more metal--poor than metal--rich stars, as can be seen in 
Figure \ref{f1}.  This means our derived metallicity distribution is affected 
by \emph{both} the actual distribution \emph{and} observational selection
effects.  Given that we only observed one star on the most metal--rich branch, 
it is possible that stars with metallicities higher than [Fe/H]=--0.75 exist. 
However, since our observed completion fraction is significantly higher for
the most metal--poor stars, it is likely that our observed distribution
function is accurate in a relative sense such that the cluster was rapidly 
enriched from the primordial metallicity of [Fe/H]$\sim$--2.15 to the first
major epoch of star formation at [Fe/H]$\sim$--1.75.  The absence of stars more 
metal--poor than [Fe/H]$\sim$--2.2 means the proto--$\omega$ Cen environment 
was already pre--enriched, perhaps from processes such as cloud--cloud 
collisions (Tsujimoto et al. 2003), when the primary metal--poor population 
formed.  In contrast, field stars in the Galactic halo exhibit a wide range of 
metallicities from [Fe/H]$>$0.0 to [Fe/H]$<$--4.0 (e.g., Gratton et al. 2004),
indicating that the two do not share a common chemical enrichment history.

\begin{figure}
\epsscale{1.00}
\plotone{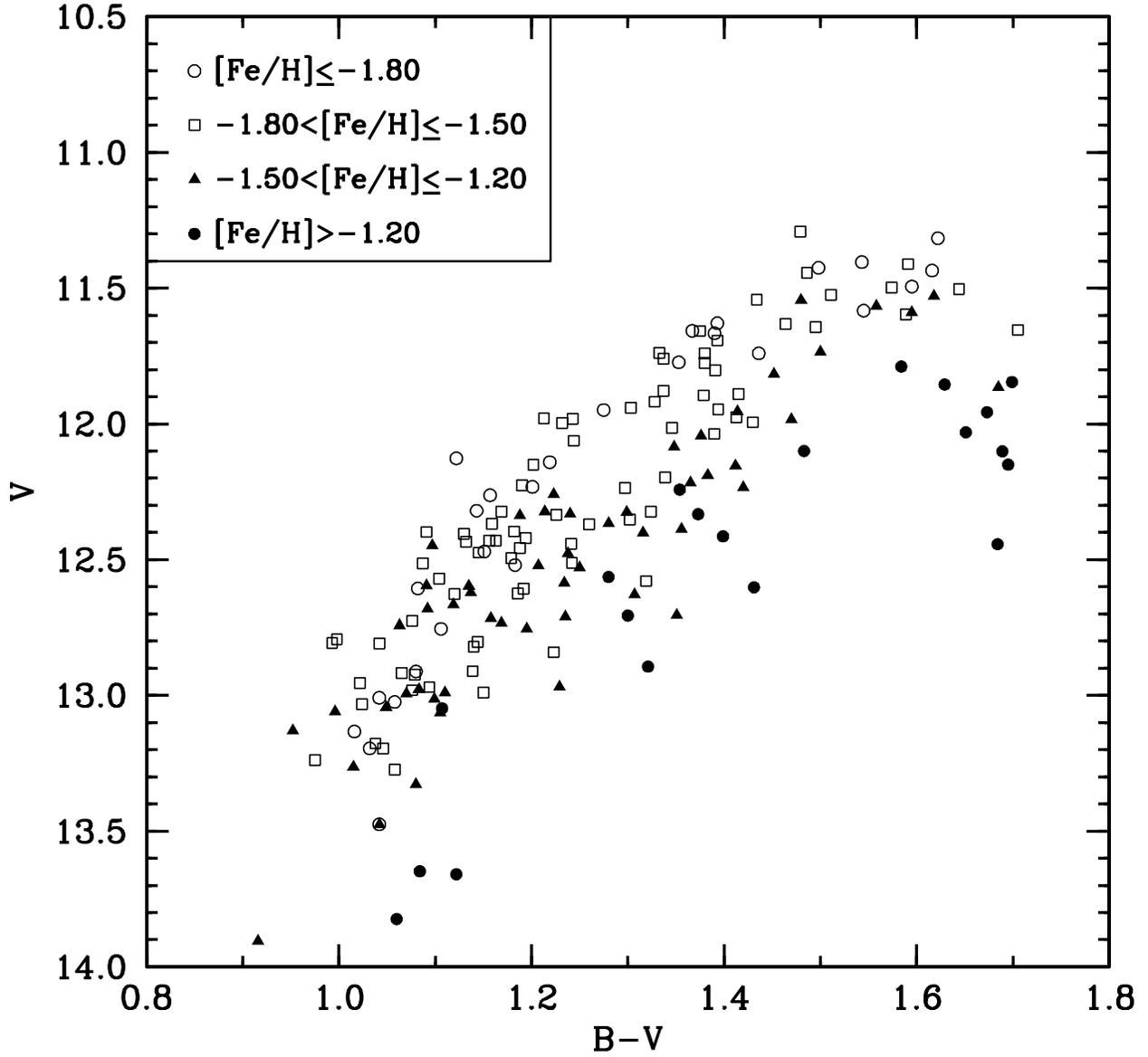}
\caption{Color--magnitude diagram of program stars displayed in various 
metallicity bins as shown above.}
\label{f11}
\end{figure}

The distribution shown in Figure \ref{f10} suggests that if $\omega$ Cen 
evolved as a single entity (i.e., without significant contributions from 
mergers), then there were four to five significant star formation episodes
that occurred.  This seems to fit the high resolution photometric data from 
Sollima et al. (2005a) and Villanova et al. (2007) that show the multiple 
giant branches appear in discrete groups instead of as a continuous 
distribution.  This trend is similarly reproduced in Figure \ref{f11}, where 
our derived metallicities are superimposed on the photometric data from van 
Leeuwen et al. (2000).  Here, even when binning by the approximate 3$\sigma$
value of each peak in the distribution from Figure \ref{f10} (0.3 dex), the 
different metallicity groups can be separated.  The metallicity distribution 
from Figure \ref{f10} is very well produced in the hydrodynamical chemical 
enrichment simulations of Marcolini et al. (2007), where they assumed $\omega$ 
Cen is the core remnant of a dwarf spheroidal galaxy that was captured and 
tidally stripped $\sim$10 Gyr ago with star formation occurring over roughly 
1.5 Gyr.  The simulated metallicity peaks from Marcolini et al. (2007) lie at 
[Fe/H]$\sim$--1.6, --1.35, --1.0, and --0.70, which are very similar to ours
at [Fe/H]=--1.75, --1.45, --1.05, and --0.75.  

There is some evidence that different metallicity populations may be spatially
and kinematically unique (Norris et al. 1996; 1997; Suntzeff \& Kraft 1996; 
Hilker \& Richtler 2000; Pancino et al. 2000; 2003).  In Figure \ref{f12}, we 
present Fe and Al abundances as a function of distance from the cluster 
center.  Keeping in mind our observational bias, we find a marginal tendency 
for the more metal--rich stars to be located in the inner regions of the 
cluster while the more metal--poor stars are rather evenly distributed at all 
radii sampled here.  However, given our small sample size in the metal--rich
regime, we are unable to make any definitive arguments for or against a 
metallicity--radius relationship.  It should be noted though that Ikuta \& 
Arimoto (2000) and Rey et al. (2004) do not find any strong evidence for the 
metal--poor and metal--rich populations having a spatially different structure.
Even though the relaxation time for $\omega$ Cen is thought to exceed 5 Gyr
(Djorgovski 1993; Merritt et al. 1997), any correlation between projected 
spatial position and metallicity is apparently subtle.  However, it has been
pointed out in deep photometric surveys (e.g., Rey et al. 2004) that the most
metal--rich RGB--a is predominately seen in CMDs of the inner region of the 
cluster.  

The main result indicating that at least the most metal--rich
population may have a different formation history is that those stars appear 
to have a lower velocity dispersion (i.e. are kinematically cooler) than the 
other populations and do not show signs of rotation (Norris et al. 1997).
In Figure \ref{f13} we show our derived radial velocities plotted both as a 
function of log $\epsilon$(Fe)\footnote{log 
$\epsilon$(X)=log(N$_{\rm X}$/N$_{\rm H}$)+12} and log $\epsilon$(Al), where 
the error bars indicate the velocity dispersion in the data.  To within one 
standard deviation, we do not find significant evidence for any of the stellar 
populations having a different bulk radial velocity or velocity dispersion.
It seems unlikely that a larger sample size would provide significantly
different results because Reijns et al. (2006) determined radial velocities
for nearly 2000 $\omega$ Cen members and concluded the RGB--a stars had radial
velocity and dispersion values consistent with the entire cluster.  Pancino
et al. (2007) have shown the rotational velocities for all populations are 
comparable to one another, but interestingly they find an underlying sinusoidal
pattern in their measured velocities as a function of position angle.  However,
the metal--poor, intermediate metallicity, and anomalous giant branches all
show the same sinusoidal pattern.  Whether any true kinematic anomaly exists 
for this cluster or not remains to be seen.  

\subsection{Al Abundances}

The bulk of aluminum production in galaxies and globular clusters is thought
to arise from quiescent carbon and neon burning in massive stars 
(M$\ga$8 M$_{\sun}$) and HBB occurring in the envelopes of IM--AGB stars via 
the MgAl cycle (e.g., Arnett \& Truran 1969; Arnett 1971).  In most Galactic 
globular clusters, there is a very small ($<$0.10 dex) spread in the abundance 
of heavy $\alpha$ and Fe--peak elements, with a somewhat larger spread 
($\sim$0.3--0.6 dex) in s-- and r--process elements (e.g., Sneden et al. 
2000).  However, the lighter elements carbon through aluminum are typically 
not uniform and in some cases show star--to--star variations of more than a 
factor of 10.  While $\omega$ Cen does not share all of the same chemical 
characteristics as globular clusters, the primary production locations of each 
element should be similar to globular clusters and/or the Galactic halo.  The 
lesson learned from the monometallicity of ``normal" globular clusters is that 
however Al manifests itself onto the surface of stars, the process must not 
alter Fe--peak, s--process, or r--process abundance ratios.   This means that 
the often large star--to--star variation of [Al/Fe] seen in globular clusters 
(but not in halo field stars) are not due to supernova yields or the 
s--process, leaving either in situ deep mixing or HBB as the possible sites 
for [Al/Fe] variation.  With these two scenarios in mind, we explore Al 
abundances with the goal of helping to constrain the source of Al variation 
and chemical evolution in $\omega$ Cen.

While the literature on Fe abundances for both evolved and main sequence stars
is quite extensive, the spectroscopic surveys by Norris \& Da Costa (1995) and 
Smith et al. (2000) represent the only studies to consider light element 
abundances that include Al for a large (N$\ge$10) number of RGB stars in 
$\omega$ Cen.  The results of those two studies indicate that the full range 
of [Al/Fe] is larger than 1.0 dex, Al and Na are correlated, Al and O are 
anticorrelated, and there is a hint of a decrease in [Al/Fe] with increasing 
[Fe/H].  We present the results of our larger sample plotting [Al/Fe] as a 
function of [Fe/H] in Figure \ref{f14}.  Even for the lowest metallicity 
stars, a large range in [Al/Fe] of $\sim$0.70 dex is already present.  Near 
the first metallicity peak at [Fe/H]=--1.75, where it is assumed the first 
episode of star formation after the initial enrichment period occurred, the 
full range in [Al/Fe] reaches a maximum value of $\sim$1.3 dex.  This 
star--to--star variation remains mostly constant until about [Fe/H]=--1.4, 
where the variation begins to decrease smoothly with increasing [Fe/H].  
Interestingly, the ``floor" Al abundance remains mostly constant at 
[Al/Fe]$\sim$+0.15, regardless of the star's metallicity; a characteristic 
shared with many globular clusters of various metallicity and in agreement 
with [Al/Fe] values typical of Galactic halo stars in $\omega$ Cen's 
metallicity regime.

In Figure \ref{f15}, we overlay a boxplot on top of the underlying 
distribution from Figure \ref{f14}.  The median [Al/Fe] ratio
typically resides between about 0.45 and 0.80 dex for all well--sampled 
metallicities, with a relatively constant interquartile range.  This implies
that the average amount of Al in the cluster must increase with increasing Fe 
abundance, at least up to [Fe/H]$\sim$--1.4.  This result is confirmed in
Figure \ref{f16}, where log $\epsilon$(Al) is plotted against log 
$\epsilon$(Fe).  It appears that for metallicities higher than about log 
$\epsilon$(Fe)=6.0 ([Fe/H]$\approx$--1.50), log $\epsilon$(Al) no longer 
increases beyond log $\epsilon$(Al)$\approx$6.40 and the star--to--star scatter
decreases.  This result is likely robust against our observational bias 
because all stars observed in the metal--rich regime are located at or near 
the RGB tip (see Figure \ref{f1}), where it is believed any Al enhancements 
due to deep mixing should be the most apparent.  However, no obvious trend is 
seen between Al abundance and evolutionary state.

\begin{figure}
\epsscale{1.00}
\plotone{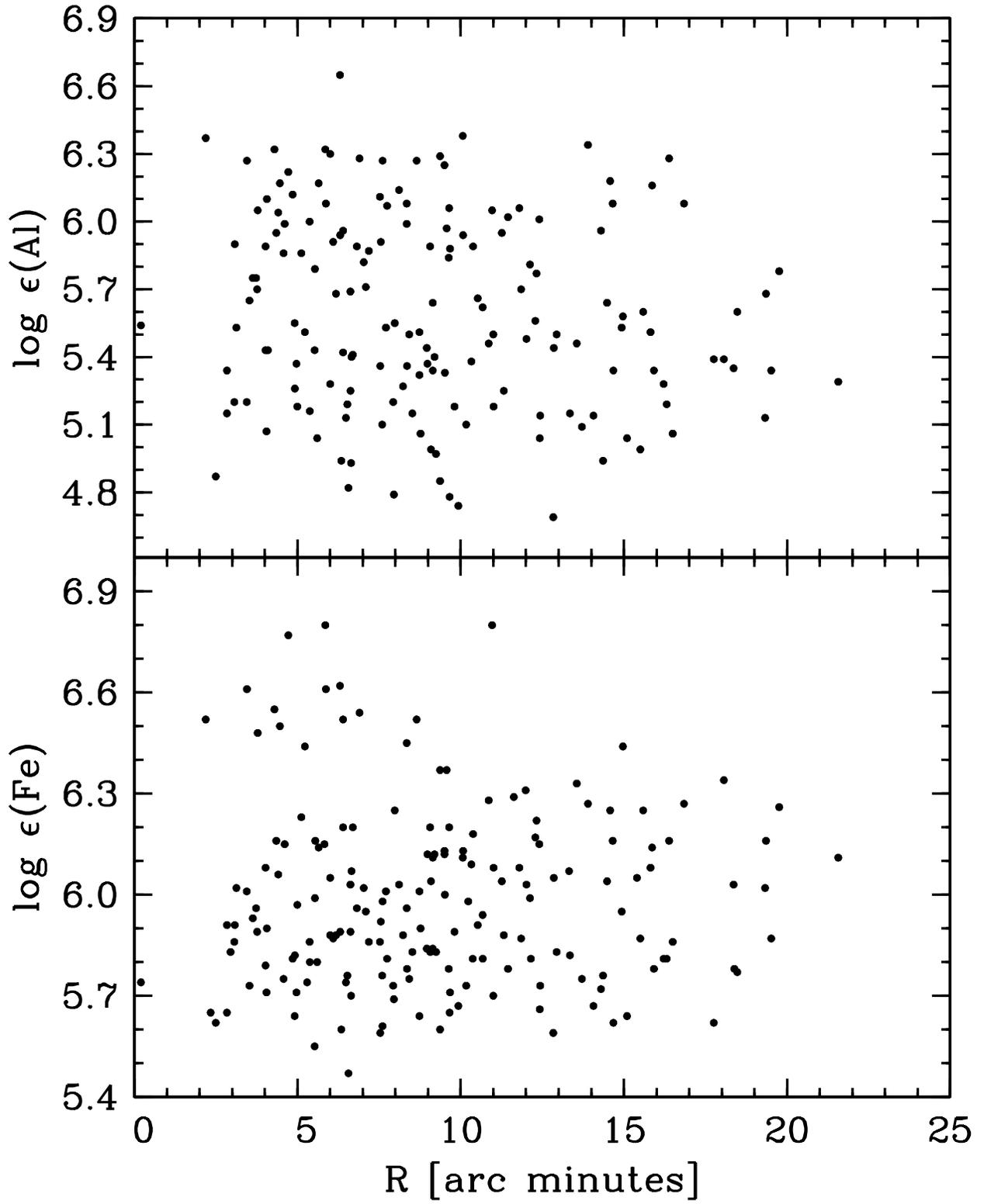}
\caption{Al and Fe are plotted as a function of radial distance from the
cluster center.}
\label{f12}
\end{figure}

\begin{figure}
\epsscale{1.00}
\plotone{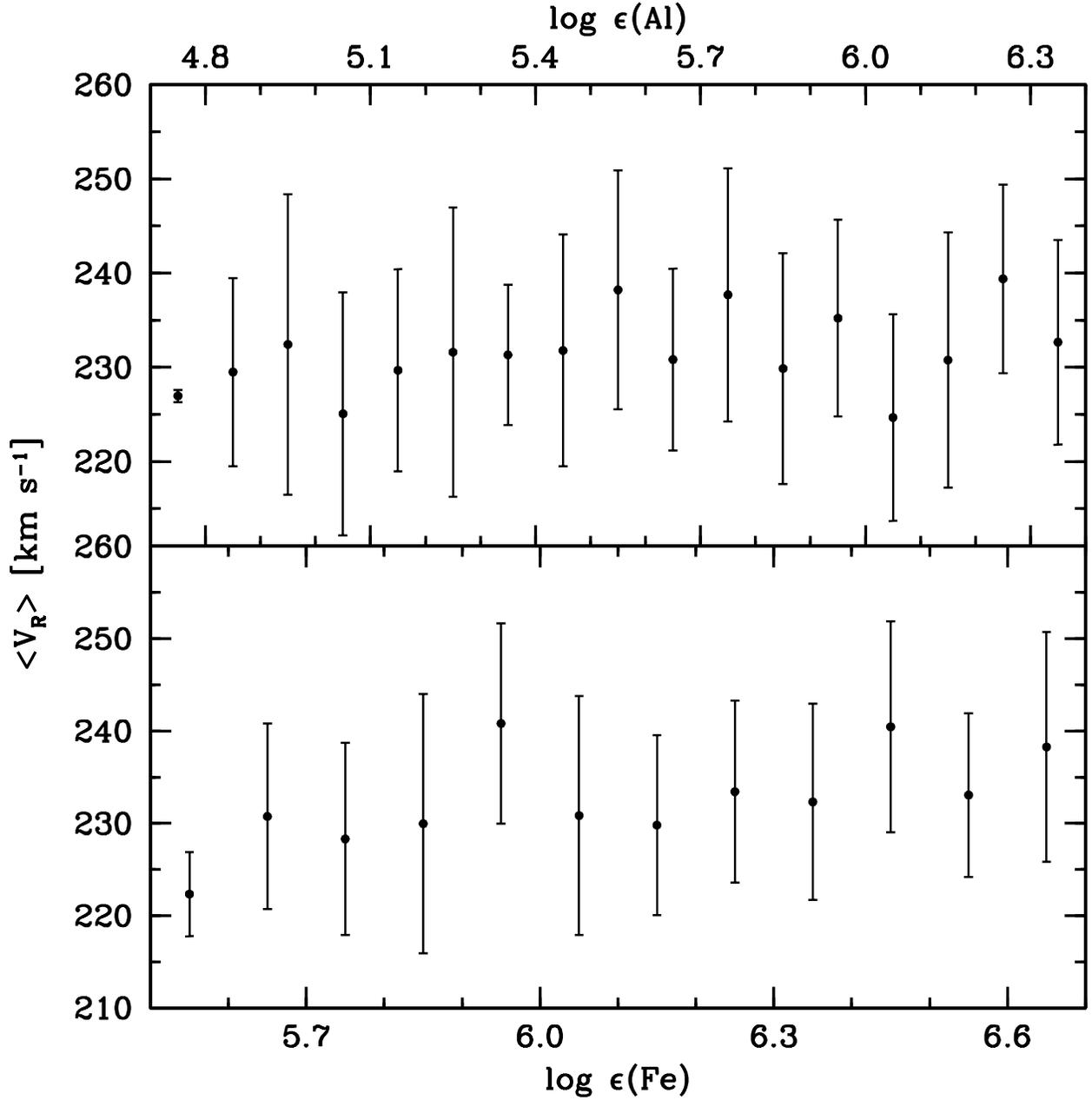}
\caption{The top panel shows average radial velocity versus log $\epsilon$(Al)
and the bottom panel is for log $\epsilon$(Fe).  The filled circles represent
average radial velocities in each abundance bin and the vertical bars indicate
the velocity dispersion in each bin.  Both panels have a bin size of 0.10 dex
in abundance.}
\label{f13}
\end{figure}

\begin{figure}
\epsscale{1.00}
\plotone{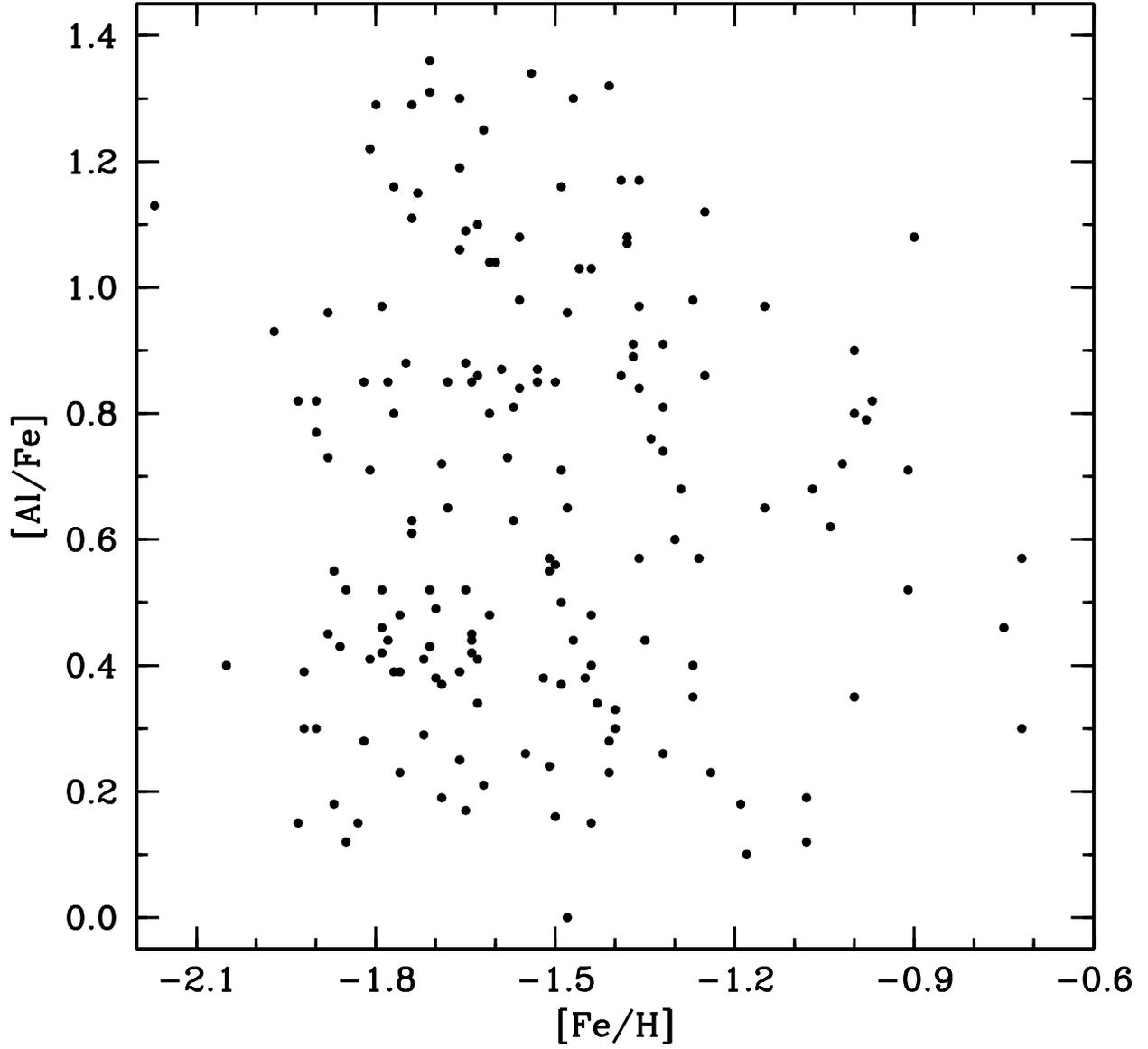}
\caption{[Al/Fe] plotted as a function of [Fe/H].}
\label{f14}
\end{figure}

\begin{figure}
\epsscale{1.00}
\plotone{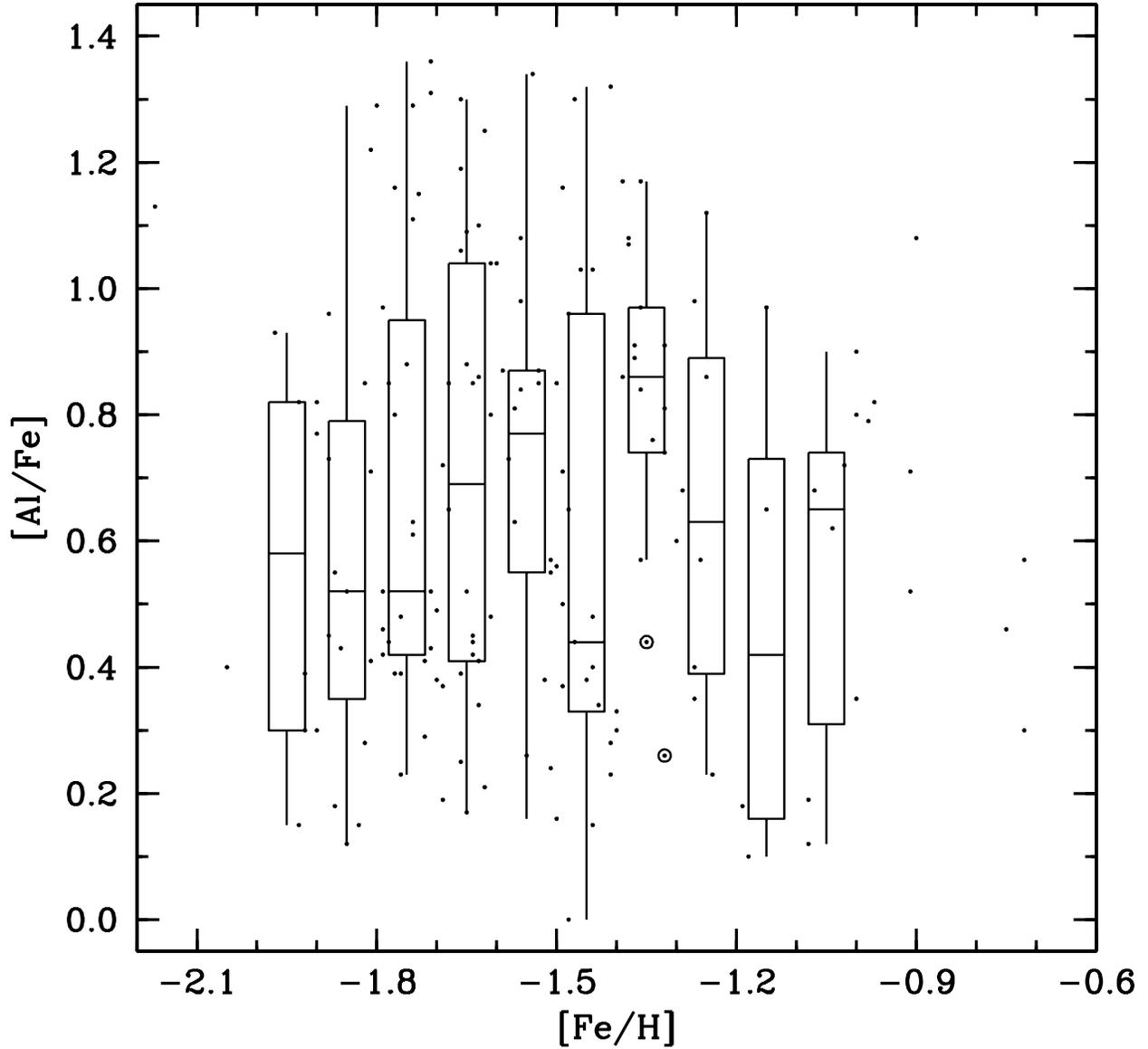}
\caption{A box plot is shown on top of the [Al/Fe] versus [Fe/H] plot given in
Figure \ref{f14}.  The data are binned into 0.10 dex intervals with the boxes
centered on each bin.  The middle line of each box indicates the median value,
the lower and upper bounds of the box are the first and third quartile, the
vertical lines are the full data range neglecting outliers, and the open 
circles indicate data lying 1.5--3.0 times the interquartile range away from
either boundary.}
\label{f15}
\end{figure}

As discussed previously, there is some evidence for a correlation between
Fe abundance and distance from the cluster center and we show the results
from this study in the bottom panel of Figure \ref{f12}.  In the top panel of
Figure \ref{f12}, we present the same data but for Al instead of Fe.  While
there may be a tendency for the most metal--rich stars to be located
inwards of about 10--15$\arcmin$, there is no evidence of a trend for Al.
Instead, stars of varying Al abundance are uniformly spread throughout the 
entire region sampled, at least out to $\sim$20$\arcmin$.  Likewise, the top 
panel of Figure \ref{f13} shows average radial velocities for Al abundances in 
0.10 dex bins.  To within uncertainties, there appears to be no trend in 
either radial velocity or velocity dispersion with log $\epsilon$(Al).  The 
fact that we do not find any preference of Al abundance or star--to--star 
dispersion with distance from the cluster center or radial velocity suggests 
star formation occurred on timescales shorter than those required to uniformly 
mix the gas.

\subsection{Possible Implications on Chemical Evolution}

From our available spectroscopic data for 180 RGB stars, we have 
confirmed the existence of at least four stellar populations ranging in 
metallicity from --2.2$<$[Fe/H]$<$--0.70, in agreement with previous
photometric, low resolution spectroscopic, and smaller sample high resolution
spectroscopic studies.  Additionally, we have determined [Al/Fe] abundances
for about 165 giants, most of which for the first time, with a sample larger
by more than a factor of four than what was previously available in the 
literature.  We find a constant Al abundance floor of [Al/Fe]$\sim$+0.15 
present at all metallicities, but with a largely varying and metallicity 
dependent spread above the floor.  The star--to--star variation reaches a
maximum extent in the intermediate metallicity regime, which is consistent with 
the second peak in the metallicity distribution, and begins to decline at 
higher metallicities.  The floor itself is consistent with observations of 
field stars and is predicted by Galactic chemical evolution models, but the 
large [Al/Fe] variations are not predicted.  Observations of some Galactic 
globular cluster stars, especially more metal--poor than [Fe/H]$\sim$--1.5,
show similar large star--to--star variations in [Al/Fe].  Combining our 
determined Fe and Al abundances with those available in the literature for 
these and other elements now allows us to examine each metallicity regime
in turn.

\subsubsection{The Metal--Poor Population}

A prominent feature of the metal--poor stars ([Fe/H]$\la$--1.6) in $\omega$ 
Cen is the rapidly increasing abundances of Na, Al, and light and heavy 
s--process elements relative to Fe as the metallicity increases from 
[Fe/H]=--2.2 to the first metallicity peak at [Fe/H]=--1.75 (e.g., Norris \& 
Da Costa 1995; Smith et al. 2000).  These increases are accompanied by nearly
constant heavy [$\alpha$/Fe]$\sim$+0.30, low Cu abundances 
([Cu/Fe]$\sim$--0.60), and low r--process abundances ([Eu/Fe]$\sim$--0.50).  
These results seem to indicate that massive stars exploding as type II SNe are 
the primary contributors for Fe--peak and heavy $\alpha$--element enhancement 
in the cluster, but the low Eu abundances, which should be synthesized in the 
same stars, are puzzling.  Additionally, the growing s--process component 
appears to be best fit by models of 1.5--3 M$_{\sun}$ AGB ejecta (Smith et al. 
2000).  The lack of clear evidence for type Ia SNe having contributed to the 
chemical composition of metal--poor stars in $\omega$ Cen (e.g., Smith et al. 
2000; Cunha et al. 2002; Pancino et al. 2002; Platais et al. 2003) is 
consistent with the $\ga$1 Gyr timescales needed for type Ia SNe to evolve and 
the fact that they might not efficiently form in metal--poor environments 
(Kobayashi et al. 1998).

\begin{figure}
\epsscale{1.00}
\plotone{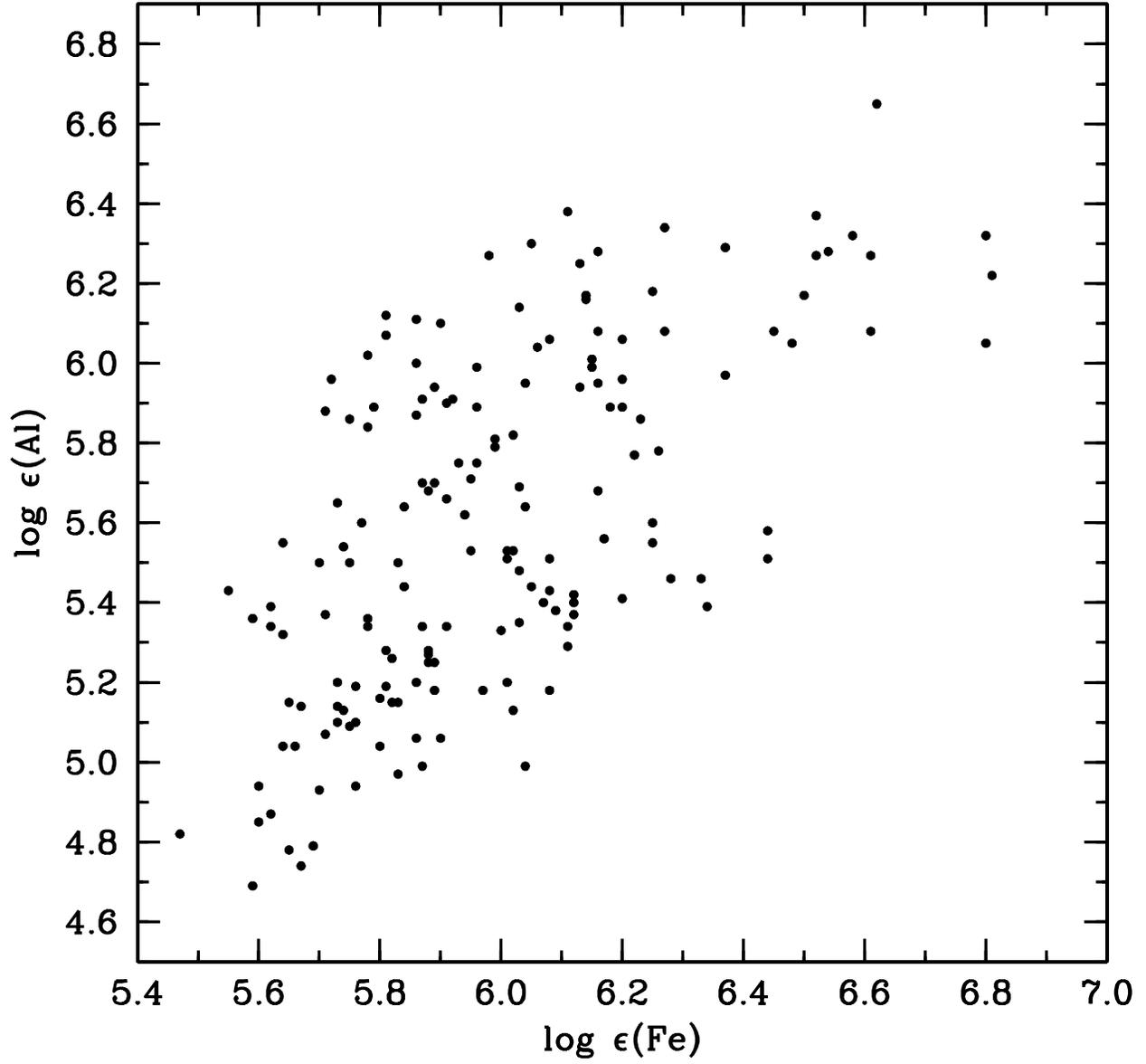}
\caption{Log $\epsilon$(Al) is plotted as a function of log $\epsilon$(Fe).}
\label{f16}
\end{figure}

As mentioned above, the majority of Al present in the atmospheres of 
these RGB stars was likely produced in type II SNe explosions that polluted the
pristine gas from which these stars formed.  While the heavy element data 
do not support high mass ($\ga$8M$_{\sun}$) stars being the source for the 
more than 1.0 dex [Al/Fe] variations, that may be explained from HBB occurring 
in IM--AGB stars, in situ deep mixing, or a hybrid scenario.  In Figures 
\ref{f14}--\ref{f16}, we have shown that [Al/Fe]$\ge$0 for \emph{all} 
metal--poor stars sampled, but a constant Al abundance floor is setup at 
[Al/Fe]$\sim$+0.15 with a rapidly increasing star--to--star dispersion that 
reaches about 1.3 dex in extent by [Fe/H]=--1.75.  For the neutron capture 
elements, which are the only other group exhibiting a variations with 
metallicity, Smith et al. (2000) showed stars with [Fe/H]$\sim$--2 are 
dominated by an r--process component with a shift to a primarily s--process 
component by [Fe/H]$\ga$--1.8.  

In the pure pollution scenario, which does not invoke deep mixing affecting 
elements heavier than N, type II SNe, low and IM--AGB stars, and perhaps winds 
from less evolved very massive stars (e.g., Maeder \& Meynet 2006) are 
responsible for all abundance anomalies.  Adding our large Al data set to the 
sample of stars previously observed may help constrain enrichment timescales 
and polluting AGB masses.  Conventional theory suggests light and s--process 
elements do not share the same origin and $\omega$ Cen's s--process component 
is best fit with lower mass AGB stars, but masses lower than $\sim$3--4 
M$_{\sun}$ undergo third dredgeup without significant HBB (e.g., Karakas \& 
Lattanzio 2007) and thus should not appreciably alter their envelope Al 
abundances.  Additionally, Ventura \& D'Antona (2007) suggest globular cluster 
light element anomalies can only be explained with ejecta from AGB stars in 
the mass range of $\sim$5--6.5 M$_{\sun}$.  While our sample only includes two 
stars with [Fe/H]$<$--2 (36036 \& 51091), the elevated [Al/Fe] ratios of +0.40 
and +1.13 suggest IM--AGB stars, with lifetimes of about 
50--150$\times$10$^{\rm 6}$ yrs (Schaller et al. 1992), have already polluted 
the $\omega$ Cen system.  In this case, the low metallicity environment would 
favor high [Al/Fe] yields from HBB processes occurring in IM--AGB stars.  The 
rapidly rising average value of log $\epsilon$(Al) shown in Figure \ref{f16} 
in the metallicity regime --2.0$\la$[Fe/H]$\la$--1.6 implies a continued 
contribution from IM--AGB stars, presumably forming from the same star 
formation event that creates the first peak in the metallicity distribution.  
The top two panels of Figure \ref{f17} show binned [Al/Fe] for this 
metallicity regime and we note approximately four sub--populations with 
[Al/Fe]$\sim$+0.15, +0.45, +0.85, and $>$+1.05.  Predicted yields from type II 
SNe (e.g., Woosley \& Weaver 1995) and measurements of field stars (e.g., 
Fulbright 2000) suggest type II SNe should enrich the ISM with 
[Al/Fe]$\sim$+0.10 to +0.30 while $\sim$5--6.5 M$_{\sun}$ AGB stars should 
produce [Al/Fe]$\sim$+0.50 to +1.10 (e.g., D'Antona \& Ventura 2007), which 
could explain our observed distribution.  Given the rather short lifetimes of 
stars believed to produce Al and the fact that evidence for 1.5--3.0 M$_{\sun}$
pollution does not appear until [Fe/H]$\sim$--1.8, it would seem that 
$\omega$ Cen was probably enriched from [Fe/H]=--2.2 to --1.75 in 
$\sim$0.5--1.0 Gyr.

\subsubsection{The Intermediate Metallicity Populations}

For the two intermediate metallicity populations ([Fe/H]=--1.45 and
[Fe/H]=--1.05), the heavy [$\alpha$/Fe] ratio remains constant and the 
s--process abundances level off with very little star--to--star dispersion
(Norris \& Da Costa 1995; Smith et al. 2000).  As in the most metal--poor 
stars, r--process and Cu ratios relative to Fe remain low and mostly 
unchanged.  However, the star--to--star scatter in O, Na, and Al is still 
quite large.  It is interesting to point out that log $\epsilon$(Al) reaches
its maximum value at about the same metallicity at which the s--process 
elements reach a constant ratio relative to Fe.  The 
[Al/Fe] abundance floor is constant throughout this metallicity regime at 
[Al/Fe]$\sim$+0.15, which means the scatter, still considerably larger than for 
[Ba/Fe], decreases as a function of increasing metallicity.  This trend should 
presumably be present for Na and in the opposite sense for O assuming the 
Na--Al correlation and O--Al anticorrelation exist at all metallicities.

Had the scatter in Al abundances been comparable to that of other heavier
elements in this metallicity range ($\sim$0.10--0.30 dex) with a nearly 
constant [Al/Fe] ratio, as is seen in field stars, we might be inclined to 
believe Al enhancement in the cluster was due solely to production in massive 
stars and that typical type II SNe ejecta have [Al/Fe]$\sim$+0.15.  It is 
interesting to note that the [Al/Fe] floor tracks closely (with a slight 
offset of $\sim$0.2-0.3 dex) to the Galactic chemical evolution model 
presented in Timmes et al. (1995; their Figure 19), assuming the amount of Fe 
ejected is decreased by a factor of two, and Samland (1998; their Figure 10), 
with an increase in secondary (i.e., metal--dependent) Al production by a 
factor of five.  If the well--known light element correlations/anticorrelations 
seen in previously observed $\omega$ Cen stars (e.g., Norris \& Da Costa 1995) 
holds at all metallicities and for all stars, those with [Al/Fe]$\sim$+0.15 
should also have [O/Fe]$\sim$+0.30, heavy [$\alpha$/Fe]$\sim$+0.30, and 
[Na/Fe]$\sim$--0.20, which are consistent with predicted yields from type II 
SNe (e.g., Woosley \& Weaver 1995).  It could be that these stars formed 
preferentially out of SNe ejecta without significant IM--AGB contamination.

\begin{figure}
\epsscale{1.00}
\plotone{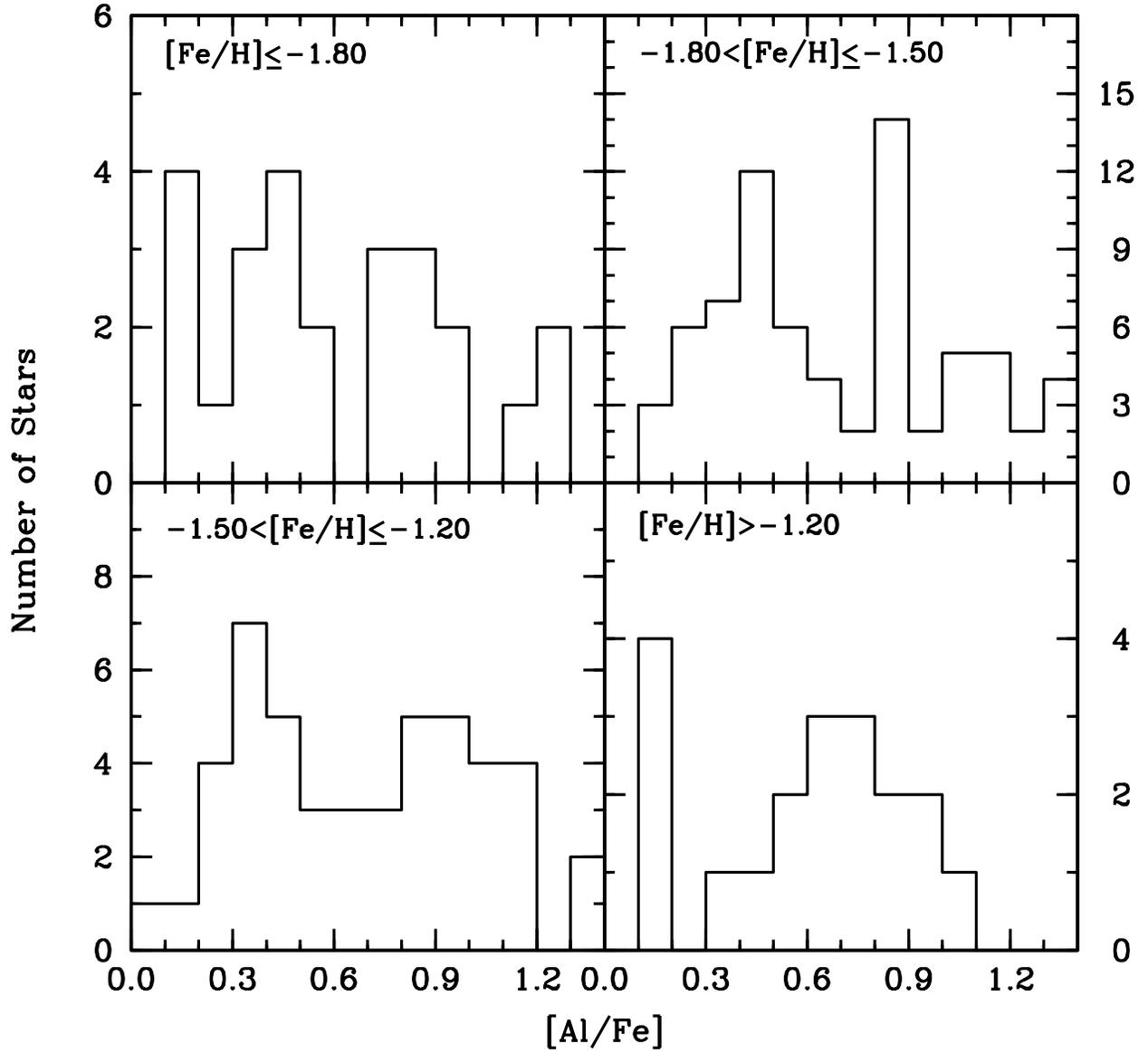}
\caption{Histograms of [Al/Fe] using a bin size of 0.10 dex for multiple
metallicity bins.}
\label{f17}
\end{figure}

While the maximum observed log $\epsilon$(Al) increases with metallicity for
the most metal--poor $\omega$ Cen giants, this trend halts at 
[Fe/H]$\sim$--1.4, which coincides with the second peak in the metallicity 
distribution (i.e., the next round of star formation).  We know the heavy
[$\alpha$/Fe], [Ba/Fe], and floor [Al/Fe] ratios remain constant at higher 
metallicities, indicating an increase in log $\epsilon$(Ba), log 
$\epsilon$($\alpha$), and the minimum log $\epsilon$(Al) that track with Fe.  
The question now posed by the Al data is why does the process producing the 
high Al values shut off or become less efficient at [Fe/H]$\ga$--1.45?  
Increases in metallicity lead to lower temperatures at the bottom of the 
convective envelope and require higher masses for HBB to occur.  It may be 
that we are observing the result of lower convective efficiency at higher 
metallicity and/or that fewer IM stars form in higher metallicity environment.
IM--AGB models in the metallicity range of --1.5$\la$[Fe/H]$\la$--0.7 (e.g., 
Fenner et al. 2004; Ventura \& D'Antona 2007; 2008) predict [Al/Fe] yields of 
$\sim$+0.5 to +1.0, with lower [Al/Fe] yields at higher [Fe/H].  This may 
explain the bimodal distribution in the bottom panels of Figure \ref{f17}, 
with the abundances in between possibly being due to varying degrees of ejecta 
dilution.  The fact that the metallicity at which the heavy elements cease to 
increase in abundance more quickly than Fe and the metallicity where the 
maximum [Al/Fe] begins to decrease coincide suggests an important parameter 
changed in $\omega$ Cen at this point in its evolution.  It may even be the 
case that this is when the progenitor dwarf galaxy began to change 
structurally via encounters with the Galactic disk.  It appears that at 
metallicities higher than [Fe/H]=--1.45, the cluster slowly approaches a 
constant [Al/Fe], which is consistent with values observed in the halo.

While type Ia ejecta have been mostly ruled out by previous studies as 
contributors to the most metal--poor population, the metallicity at which they
become important contributors is unclear.  Marcolini et al. (2007) claim that 
their intermediate metallicity peak at [Fe/H]$\sim$--1.4 is due primarily to 
inhomogeneous pollution by type Ia SNe.  It is interesting to note that in this
same metallicity bin we find a median [Al/Fe] value about 0.40 dex lower than
the two surrounding bins as well as the only star with [Al/Fe]$\la$+0.15.  It is
uncertain whether this is a real effect or simply due to an anomalous selection
of stars.  Inhomogeneous pollution by type Ia SNe may also explain the bimodal
distribution seen in the bottom panels of Figure \ref{f17} where stars
polluted by both type Ia ejecta and IM--AGB stars exhibit lower [Al/Fe] ratios
and ``normal" stars polluted by type II SNe and IM--AGB stars have higher 
[Al/Fe] values.  While the same trend is not particularly apparent for
s--process elements (e.g., Smith et al. 2000), this may be due to a smaller 
sample size, especially if inhomogeneous pollution only affected a small 
percentage of intermediate metallicity stars; however, this could explain the
few observations in the literature of stars with [Fe/H]$\sim$--1.4 and 
[Ba/Fe]$\sim$0 (e.g., Smith et al. 1995).

\subsubsection{The Metal--Rich Population}

For stars more metal--rich than [Fe/H]$\sim$--1, there is some evidence of
a decrease in [$\alpha$/Fe] and an increase in [Cu/Fe] (Pancino et al. 2002;
but see also Cunha et al. 2002), which, if true, likely indicates an increased
contribution from type Ia SNe.  Similarly, there appears to be a decrease in
[Eu/Fe] with perhaps a similar decrease in the abundance of s--process elements
relative to Fe (Norris \& Da Costa; Smith et al. 2000).  Although the Al data 
are rather incomplete in this metallicity regime, the general trends seen in 
slightly more metal--poor stars appear to continue.

While the scope of an age spread amongst the various metallicity populations
is still unknown, the Al data presented here seem to indicate that the age
difference between the intermediate and metal--rich populations is not
especially large.  In particular, stars with the largest values of log
$\epsilon$(Al) appear with [Fe/H] ranging from --1.5 to --0.7, perhaps 
indicating that they formed from gas polluted by the same generation of 
IM--AGB ejecta.  In this scenario, the lower [Al/Fe] ratios at high metallicity
might be due to those stars forming in regions where [Fe/H] increased due to
inhomogeneous pollution by type Ia SNe, as mentioned in Marcolini et al. 
(2007).  In their scenario, this effect should be more important for the inner
regions of the cluster.  This may be corroborated by our finding that there
is no apparent relationship between log $\epsilon$(Al) and distance from the
cluster center, but a trend might be present for Fe such that stars with 
[Fe/H]$>$--1 are preferentially located closer to the cluster center.  In any
case, additional data are required in this metallicity regime to determine 
whether the decreasing [Al/Fe] ratios are a real effect or the result of 
incomplete statistics.  It will be interesting to see if O and Na display 
similar behavior to Al as a function of [Fe/H].

\section{SUMMARY}

We have determined radial velocities, Fe, and Al abundances for 180 RGB stars 
in the Galactic globular cluster $\omega$ Cen using moderate resolution
(R$\approx$13,000) spectroscopy.  The bulk of our sample includes stars with 
V$<$14.0, but an observational bias is present such that we preferentially
observed more luminous and more metal--poor stars.  The spectra ranged from
6500--6750 \AA\ and Fe abundances were based on an average of approximately
10--20 Fe I lines.  Al abundances were determined using either spectrum 
synthesis or equivalent width analyses of the 6696, 6698 \AA\ Al I doublet,
with synthesis being reserved for CN--strong and/or metal--rich stars.

With respect to our determined Fe abundances, we find in agreement with 
previous studies that at least four or more different metallicity populations 
are present in the cluster.  Peaks in the metallicity distribution function 
appear at [Fe/H]=--1.75, --1.45, --1.05, and --0.75, indicating the presence of
multiple star formation episodes.  We do not find evidence suggesting any of 
the different metallicity populations are kinematically or spatially unique,
but it should be noted that our observed completion fraction is low for stars
more metal--rich than [Fe/H]$\sim$--1.0 and we only observed stars between about
2$\arcmin$ and 20$\arcmin$ from the cluster center.

Our Al data corroborate the Fe results such that there does not appear to be 
any correlation between Al abundance and distance from the cluster center or
radial velocity.  This suggests that the cluster gas was not significantly 
mixed while star formation was still occurring.  In a plot of [Al/Fe] versus 
[Fe/H], the data reveal a star--to--star variation of nearly 1.3 dex that 
stays mostly constant until [Fe/H]$\sim$--1.45, in which case the spread in 
[Al/Fe] declines monotonically with increasing [Fe/H].  Additionally, the 
[Al/Fe] floor remains nearly constant across all metallicities sampled here at 
[Al/Fe]$\sim$+0.15.  This result is similar to what is predicted based on type 
II SNe yields and closely mimics the trend seen in Galactic halo field stars. 
The anomalously low median [Al/Fe] ratio at [Fe/H]=--1.45 may be evidence 
for inhomogeneous pollution from type Ia SNe and could explain the bimodal
[Al/Fe] distribution seen in intermediate metallicity stars, but more 
observations are required to confirm whether this is real or the result of 
an incomplete sample.

The source of the [Al/Fe] spread that has also been observed in other light 
elements remains an open problem, but the results obtained here pose some 
interesting questions.  A plot of log $\epsilon$(Al) versus log $\epsilon$(Fe) 
shows that log $\epsilon$(Al) no longer increases beyond about 6.40 at 
metallicities higher than [Fe/H]$\sim$--1.45, which is coincident with the 
second peak in the metallicity distribution function.  Apparently, whatever 
process is responsible for manifesting very high Al abundances shuts down or 
becomes less efficient at intermediate and high metallicities.  In ``normal" 
metal--poor globular clusters, the large star--to--star variations seen in the
light elements are not shared by Fe--peak and neutron capture elements, and
it has been suggested that HBB occurring in IM--AGB stars or in situ deep 
mixing are responsible for the light element abundance anomalies.  Without
a comparable sample of O and Na data to supplement the Al abundances here,
it is difficult to determine the role either source plays.  However, AGB yields
of stars undergoing HBB indicate stars forming from material polluted by
AGB ejecta can only reach [Al/Fe] ratios between about +0.5 and +1.0, with
perhaps slightly lower and higher values being reached in higher and lower
metallicity environments, respectively.

It may be possible to explain the Al data such that core--collapse SNe drive 
the [Al/Fe] floor and an AGB mass spectrum with varying HBB efficiencies and 
mixing depths are responsible for much of the additional scatter present.  The 
decrease in the maximum [Al/Fe] with increasing [Fe/H] might then be 
attributed to requiring higher mass stars for HBB to occur at temperatures 
adequate to activate the full \iso{24}{Mg} to \iso{27}{Al} cycle, which means 
the burning material is exposed for a shorter amount of time and thus leads to 
less [Al/Fe] enhancement.  Whether this can be made to work quantitatively in 
light of the problems associated with AGB pollution scenarios (see $\S$1) 
remains to be seen.

\acknowledgments

We would like to thank the anonymous referee for a detailed and helpful report
which improved the manuscript and for pointing out the possible significance 
of type Ia SN pollution at intermediate metallicities.  We would also like to 
thank Bob Kraft and Chris Sneden for helpful discussions regarding this paper 
and Bertrand Plez for providing an electronic copy of his CN linelist.  This 
publication makes use of data products from the Two Micron All Sky Survey, 
which is a joint project of the University of Massachusetts and the Infrared 
Processing and Analysis Center/California Institute of Technology, funded by 
the National Aeronautics and Space Administration and the National Science 
Foundation. This research has made use of NASA's Astrophysics Data System 
Bibliographic Services.  This research has made use of the SIMBAD database, 
operated at CDS, Strasbourg, France.  Support for DS was provided by grant 
AST-0139617 from the NSF for a summer REU program.  Support of the College of 
Arts and Sciences and the Daniel Kirkwood fund at Indiana University 
Bloomington for CIJ is gratefully acknowledged.

{\it Facilities:} \facility{CTIO}

\clearpage

\tablenum{1}
\tablecolumns{4}
\tablewidth{0pt}

\begin{deluxetable}{cccc}
\tablecaption{Hydra Observations of $\omega$ Cen Giants}
\tablehead{
\colhead{Hydra}	&
\colhead{Wavelength}	&
\colhead{UT Date}	&
\colhead{Exposure}	\\
\colhead{Setup}	&
\colhead{[\AA]}	&
\colhead{}	&
\colhead{[s]}
}

\startdata

1	&	6600	&	2003 July 17	&	1 x 1800 \\
2	&	6600	&	2003 July 18	&	1 x 1800 \\
2	&	6600	&	2003 July 18	&	4 x 2700 \\
3	&	6600	&	2003 July 19	&	2 x 2700 \\
3	&	6600	&	2003 July 19	&	1 x 3600 \\

\enddata

\end{deluxetable}

\clearpage

\tablenum{2}
\tablecolumns{6}
\tablewidth{0pt}

\begin{deluxetable}{cccccc}
\tablecaption{Radial Velocity and Membership Information}
\tablehead{
\colhead{Star\tablenotemark{a}}	&
\colhead{Alt. ID\tablenotemark{b}}	&
\colhead{V$_{\rm R}$}	&
\colhead{Error}	&
\colhead{$\sigma$ from Mean}	&
\colhead{Mem. Prob.\tablenotemark{c}}	\\
\colhead{}	&
\colhead{}      &
\colhead{[km s$^{\rm -1}$]}      &
\colhead{[km s$^{\rm -1}$]}      &
\colhead{}      &
\colhead{}
}

\startdata
9	&	370	&	211.9	&	1.8	&	1.7	&	99	\\
5009	&	548	&	230.7	&	1.6	&	0.1	&	100	\\
6017	&	240	&	248.4	&	1.6	&	1.4	&	98	\\
10012	&	43	&	236.2	&	1.6	&	0.4	&	98	\\
11019	&	537	&	238.3	&	1.7	&	0.6	&	99	\\
11024	&	91	&	221.7	&	1.3	&	0.9	&	99	\\
12013	&	394	&	230.4	&	1.4	&	0.1	&	98	\\
14010	&	435	&	245.5	&	3.3	&	1.2	&	98	\\
\enddata

\tablenotetext{a}{Identifier from van Leeuwen et al. (2000).}
\tablenotetext{b}{Identifier from Woolley (1966).}
\tablenotetext{c}{Membership probability from van Leeuwen et al. (2000).}

\end{deluxetable}

\clearpage

\tablenum{3}
\tablecolumns{10}
\tablewidth{0pt}

\begin{deluxetable}{cccccccccc}
\tablecaption{Photometry and Model Atmosphere Parameters}
\tablehead{
\colhead{Star\tablenotemark{a}}	&
\colhead{Alt. ID\tablenotemark{b}}	&
\colhead{V}	&
\colhead{B--V}	&
\colhead{V--K}	&
\colhead{M$_{\rm V}^{\rm 0}$}	&
\colhead{\emph{T$_{\rm eff}$}}	&
\colhead{\emph{log g}}	&
\colhead{[Fe/H]}	&
\colhead{\emph{V$_{\rm t}$}}	\\
\colhead{} &
\colhead{}      &
\colhead{}     &
\colhead{}  &
\colhead{TCS} &
\colhead{} &
\colhead{[K]} &
\colhead{[cm s$^{\rm -2}$]} &
\colhead{Spec.}       &
\colhead{[km s$^{\rm -1}$]}
}
\startdata
9	&	370	&	12.529	&	1.250	&	2.870	&	$-$1.543	&	4460	&	1.20	&	$-$1.26	&	1.95	\\
5009	&	548	&	12.912	&	1.080	&	2.841	&	$-$1.160	&	4525	&	1.40	&	$-$1.90	&	1.60	\\
6017	&	240	&	12.233	&	1.420	&	3.387	&	$-$1.839	&	4110	&	0.85	&	$-$1.36	&	1.85	\\
10012	&	43	&	11.529	&	1.618	&	3.782	&	$-$2.543	&	3900	&	0.40	&	$-$1.49	&	2.10	\\
11019	&	537	&	12.841	&	1.223	&	2.985	&	$-$1.231	&	4450	&	1.30	&	$-$1.57	&	2.00	\\
11024	&	91	&	11.738	&	1.333	&	3.291	&	$-$2.334	&	4200	&	0.70	&	$-$1.76	&	1.90	\\
12013	&	394	&	12.579	&	1.319	&	3.142	&	$-$1.493	&	4275	&	1.10	&	$-$1.50	&	2.05	\\
14010	&	435	&	12.807	&	0.993	&	2.647	&	$-$1.265	&	4635	&	1.45	&	$-$1.74	&	1.40	\\
\enddata

\tablenotetext{a}{Identifier from van Leeuwen et al. (2000).}
\tablenotetext{b}{Identifier from Woolley (1966).}

\end{deluxetable}

\clearpage

\tablenum{4}
\tablecolumns{4}
\tablewidth{0pt}

\begin{deluxetable}{cccc}
\tablecaption{Line list}
\tablehead{
\colhead{Element}	&
\colhead{$\lambda$}	&
\colhead{Exc. Pot.}	& 
\colhead{log gf} \\
\colhead{}	&
\colhead{[\AA]}     &
\colhead{[eV]}     &
\colhead{}
}
\startdata
Fe II	&	6516.08	&	2.89	&	$-$3.45	\\
Fe I	&	6533.93	&	4.56	&	$-$1.36	\\
Fe I	&	6546.24	&	2.76	&	$-$1.54	\\
Fe I	&	6551.68	&	0.99	&	$-$5.77	\\
Fe I	&	6574.25	&	0.99	&	$-$5.02	\\
Fe I	&	6592.92	&	2.73	&	$-$1.47	\\
Fe I	&	6593.88	&	2.43	&	$-$2.42	\\
Fe I	&	6597.57	&	4.79	&	$-$0.95	\\
Fe I	&	6608.04	&	2.28	&	$-$3.96	\\
Fe I	&	6609.12	&	2.56	&	$-$2.69	\\
Fe I	&	6625.02	&	1.01	&	$-$5.37	\\
Fe I	&	6627.54	&	4.55	&	$-$1.58	\\
Fe I	&	6633.75	&	4.79	&	$-$0.80	\\
Fe I	&	6646.96	&	2.61	&	$-$3.96	\\
Fe I	&	6648.12	&	1.01	&	$-$5.92	\\
Fe I	&	6677.99	&	2.69	&	$-$1.35	\\
Al I	&	6696.03	&	3.14	&	$-$1.57	\\
Al I	&	6698.66	&	3.14	&	$-$1.89	\\
Fe I	&	6703.57	&	2.76	&	$-$3.01	\\
Fe I	&	6710.32	&	1.48	&	$-$4.83	\\
Fe I	&	6726.67	&	4.61	&	$-$1.07	\\
Fe I	&	6733.15	&	4.64	&	$-$1.48	\\
Fe I	&	6739.52	&	1.56	&	$-$4.79	\\
\enddata

\end{deluxetable}

\clearpage
\pagestyle{empty}
\setlength{\voffset}{25mm}

\tablenum{5}
\tablecolumns{24}
\tablewidth{0pt}

\begin{deluxetable}{cccccccccccccccccccccccc}
\tabletypesize{\scriptsize}
\rotate
\tablecaption{Equivalent Widths\tablenotemark{a,b}}
\tablehead{
\colhead{$\lambda$\tablenotemark{c}}	&
\colhead{6516}	&
\colhead{6533}     &
\colhead{6546}     &
\colhead{6551}     &
\colhead{6574}     &
\colhead{6592}     &
\colhead{6593}     &
\colhead{6597}     &
\colhead{6608}     &
\colhead{6609}     &
\colhead{6625}     &
\colhead{6627}     &
\colhead{6633}     &
\colhead{6646}     &
\colhead{6648}     &
\colhead{6677}     &
\colhead{6696}     &
\colhead{6698}     &
\colhead{6703}     &
\colhead{6710}     &
\colhead{6726}     &
\colhead{6733}     &
\colhead{6739}
}
\startdata
9\tablenotemark{d}	&	63	&	20	&	151	&	\nodata	&	\nodata	&	159	&	130	&	34	&	40	&	100	&	51	&	14	&	\nodata	&	15	&	36	&	170	&	\nodata	&	23	&	\nodata	&	\nodata	&	28	&	\nodata	&	49	\\
5009	&	\nodata	&	9	&	92	&	6	&	25	&	115	&	62	&	11	&	\nodata	&	31	&	14	&	\nodata	&	\nodata	&	\nodata	&	8	&	105	&	\nodata	&	9	&	18	&	10	&	6	&	\nodata	&	7	\\
6017	&	41	&	30	&	140	&	\nodata	&	108	&	179	&	134	&	25	&	52	&	102	&	105	&	16	&	\nodata	&	39	&	82	&	202	&	Syn	&	Syn	&	91	&	81	&	31	&	25	&	49	\\
10012	&	54	&	21	&	179	&	62	&	133	&	186	&	157	&	27	&	77	&	145	&	116	&	20	&	\nodata	&	\nodata	&	73	&	210	&	Syn	&	Syn	&	83	&	95	&	35	&	\nodata	&	\nodata	\\
11019	&	\nodata	&	13	&	139	&	18	&	68	&	141	&	118	&	17	&	17	&	82	&	24	&	8	&	\nodata	&	\nodata	&	15	&	141	&	22	&	18	&	\nodata	&	22	&	24	&	\nodata	&	33	\\
11024	&	62	&	11	&	143	&	24	&	82	&	146	&	120	&	18	&	43	&	94	&	49	&	7	&	\nodata	&	20	&	27	&	158	&	11	&	\nodata	&	38	&	42	&	18	&	9	&	35	\\
12013	&	\nodata	&	18	&	149	&	31	&	89	&	165	&	145	&	22	&	37	&	\nodata	&	74	&	17	&	\nodata	&	\nodata	&	\nodata	&	161	&	Syn	&	Syn	&	61	&	64	&	\nodata	&	9	&	\nodata	\\
14010	&	\nodata	&	\nodata	&	80	&	\nodata	&	24	&	86	&	67	&	\nodata	&	14	&	42	&	13	&	\nodata	&	\nodata	&	\nodata	&	\nodata	&	109	&	\nodata	&	\nodata	&	19	&	10	&	\nodata	&	\nodata	&	\nodata	\\
\enddata

\tablenotetext{a}{The designation ``Syn" indicates a synthetic spectrum
comparison method was used.}
\tablenotetext{b}{Equivalent widths are given in units of m\AA.}
\tablenotetext{c}{Wavelengths are given in units of \AA.}
\tablenotetext{d}{Designation is from van Leeuwen et al. (2000).}

\end{deluxetable}

\clearpage
\pagestyle{plaintop}
\setlength{\voffset}{0mm}

\tablenum{6}
\tablecolumns{9}
\tablewidth{0pt}

\begin{deluxetable}{ccccccccc}
\tablecaption{Derived Abundances}
\tablehead{
\colhead{Star\tablenotemark{a}}	&
\colhead{log $\epsilon$(Fe)}      &
\colhead{[Fe/H]\tablenotemark{b}}      &
\colhead{Num. Lines}      &
\colhead{$\sigma$/$\sqrt{N}$}      &
\colhead{log $\epsilon$(Al)}      &
\colhead{[Al/Fe]\tablenotemark{c}}      &
\colhead{Num. Lines}      &
\colhead{$\sigma$/$\sqrt{N}$}
}

\startdata
9	&	6.26	&	$-$1.26	&	14	&	0.03	&	5.78	&	0.57	&	1	&	\nodata	\\
5009	&	5.62	&	$-$1.90	&	15	&	0.05	&	5.39	&	0.82	&	1	&	\nodata	\\
6017	&	6.16	&	$-$1.36	&	18	&	0.05	&	6.28	&	1.17	&	2	&	0.07	\\
10012	&	6.03	&	$-$1.49	&	16	&	0.03	&	5.35	&	0.37	&	2	&	0.02	\\
11019	&	5.95	&	$-$1.57	&	16	&	0.03	&	5.53	&	0.63	&	2	&	0.11	\\
11024	&	5.76	&	$-$1.76	&	19	&	0.03	&	4.94	&	0.23	&	1	&	\nodata	\\
12013	&	6.02	&	$-$1.50	&	14	&	0.03	&	5.13	&	0.16	&	2	&	0.07	\\
14010	&	5.78	&	$-$1.74	&	10	&	0.03	&	\nodata	&	\nodata	&	\nodata	&	\nodata	\\
\enddata

\tablenotetext{a}{Identifier from van Leeuwen et al. (2000).}
\tablenotetext{b}{Assumed the solar log $\epsilon$(Fe)=7.52 (Sneden et al. 
1991).}
\tablenotetext{c}{Assumed the solar log $\epsilon$(Al)=6.47 (Anders \& 
Grevesse 1989).}

\end{deluxetable}

\clearpage

\tablenum{7}
\tablecolumns{5}
\tablewidth{0pt}

\begin{deluxetable}{ccccc}
\tablecaption{Abundance Sensitivity to Model Parameters}
\tablehead{
\colhead{Element}	&
\colhead{$\Delta$\emph{T$_{\rm eff}$}$\pm$100}	&
\colhead{$\Delta$\emph{log g}$\pm$0.25}	&
\colhead{$\Delta$\emph{V$_{\rm t}$}$\pm$0.25}	&
\colhead{$\Delta$N$\pm$0.30}	\\
\colhead{}       &
\colhead{[K]} &
\colhead{[cm s$^{\rm -2}$]}        &
\colhead{[km s$^{\rm -1}$]}	&
\colhead{[dex]}
}
\startdata
\hline
\multicolumn{5}{c}{[Fe/H]$\approx$--2.0}    \\
\hline
Fe I	&	$\pm$0.17	&	$\mp$0.02	&	$\mp$0.04	&	\nodata	\\
Fe II	&	$\mp$0.05	&	$\pm$0.11	&	$\mp$0.03	&	\nodata	\\
Al I	&	$\pm$0.07	&	$\mp$0.02	&	$\pm$0.00	&	$\mp$0.02	\\
\hline
\multicolumn{5}{c}{[Fe/H]$\approx$--1.5}    \\
\hline
Fe I    &       $\pm$0.16       &       $\pm$0.00       &       $\mp$0.06 	&	\nodata	\\
Fe II   &       $\mp$0.06       &       $\pm$0.12       &       $\mp$0.04	&	\nodata \\
Al I    &       $\pm$0.09       &       $\mp$0.01       &       $\mp$0.01	&	$\mp$0.05 \\
\hline
\multicolumn{5}{c}{[Fe/H]$\approx$--1.0}    \\
\hline
Fe I    &       $\pm$0.10       &       $\pm$0.01       &       $\mp$0.08 	&	\nodata	\\
Fe II   &       $\mp$0.08       &       $\pm$0.11       &       $\mp$0.04 	&	\nodata	\\
Al I    &       $\pm$0.08       &       $\pm$0.00       &       $\mp$0.01	&	$\mp$0.08 \\
\enddata

\end{deluxetable}

\clearpage

\clearpage

\clearpage

\clearpage

\clearpage

\clearpage

\clearpage

\clearpage

\clearpage

\clearpage

\clearpage

\clearpage

\clearpage

\clearpage

\clearpage

\clearpage

\clearpage

\end{document}